\documentclass[10pt,journal,compsoc]{IEEEtran}
\ifCLASSOPTIONcompsoc
  \usepackage[nocompress]{cite}
\else
  \usepackage{cite}
\fi
\ifCLASSINFOpdf
\else
\fi
\usepackage{amsmath,amssymb,amsfonts}
\usepackage{algorithmic}
\usepackage[linesnumbered,boxed,ruled,commentsnumbered]{algorithm2e}
\usepackage{subfigure}
\usepackage{multirow}
\usepackage{caption}
\usepackage{balance}
\usepackage{float}
\usepackage{booktabs}
\usepackage{graphicx}
\usepackage{textcomp}
\usepackage{xcolor}
\usepackage{color}
\usepackage{soul}
\usepackage{enumitem}

\newcommand{\model}{IMA~}
\newcommand{\modelns}{IMA}
\newcommand{\modelb}{EMA~}
\newcommand{\modelnsb}{EMA}
\newtheorem{definition}{Definition}
\newcommand{\eat}[1]{}

\begin{document}
%
\title{A Framework for Elastic Adaptation of User Multiple Intents in Sequential Recommendation}
%
%
%
%

\author{Zhikai~Wang
        and~Yanyan~Shen,~\IEEEmembership{Member,~IEEE,}}
\markboth{Journal of \LaTeX\ Class Files,~Vol.~14, No.~8, August~2015}%
{Shell \MakeLowercase{\textit{et al.}}: Bare Demo of IEEEtran.cls for Computer Society Journals}
%



\IEEEtitleabstractindextext{%
\begin{abstract}
Recently, substantial research has been conducted on sequential recommendation, with the objective of forecasting the subsequent item by leveraging a user's historical sequence of interacted items. Prior studies employ both capsule networks and self-attention techniques to effectively capture diverse underlying intents within a user's interaction sequence, thereby achieving the most advanced performance in sequential recommendation. However, users could potentially form novel intents from fresh interactions as the lengths of user interaction sequences grow. Consequently, models need to be continually updated or even extended to adeptly encompass these emerging user intents, referred as incremental multi-intent sequential recommendation.
In this paper, we propose an effective \textbf{I}ncremental learning framework for user \textbf{M}ulti-intent \textbf{A}daptation in sequential recommendation called IMA, which augments the traditional fine-tuning strategy with the existing-intents retainer, new-intents detector, and projection-based intents trimmer to adaptively expand the model to accommodate user's new intents and prevent it from forgetting user's existing intents. Furthermore, we upgrade the \model into an \textbf{E}lastic \textbf{M}ulti-intent \textbf{A}daptation~(\modelnsb) framework which can elastically remove inactive intents and compress user intent vectors under memory space limit.
Extensive experiments on real-world datasets verify the effectiveness of the proposed \model and \modelb on incremental multi-intent sequential recommendation, compared with various baselines.
\end{abstract}

}

\maketitle

\IEEEdisplaynontitleabstractindextext

%
\IEEEpeerreviewmaketitle

\IEEEraisesectionheading{\section{Introduction}\label{sec:introduction}}

%
%
%
%
\IEEEPARstart{S}{equential} recommendation, which aims at predicting user preference on items given historical user interaction sequence, is an essential task in nowadays recommender systems.
Various approaches have been proposed for sequential recommendation~\cite{icde2,caser,dien,GRU4REC}. They typically extract one user preference vector from the interaction sequence. 
Recently, many works\cite{MIND,ComiRec,limarec,FeSAIL,IMSR} argued that users usually have multiple latent intents beneath their interaction sequences, and hence they proposed to incorporate different kinds of multi-intent extractors into sequential recommendation models, which is referred to as multi-intent sequential recommendation (MSR). Specifically, existing works employ the dynamic routing or self-attention mechanism to explicitly compute the user's multiple intent vectors based on sequentially interacted items. The final user preference vector is derived by aggregating multiple intent vectors and used for next item prediction.

In real recommender systems, the lengths of user interaction sequences are ever-increasing, and users might develop new intents from new interactions. 
Both existing and new intents might reappear in future interactions. As stated in the experiments of the previous work~\cite{mimn,TAMIC}, over eighty percent of intents will reappear for more than three times. However, we do not know which of the existing or new intents will reappear.
Therefore, it is of great importance to update an MSR model continuously to capture users' new intents while retaining all existing intents. 

A simple strategy for model updating is to retrain the model using the whole user interaction sequences per time span~(e.g., one hour or a day). In this strategy, both existing and newly developed intents in the user interaction sequence can be captured. However, training on the whole user interaction sequences is extremely time-consuming, and this strategy is not practical in situations where all the historical interactions are not obtainable or the memory space for maintaining the whole sequences during training and inference is constrained.
To address this problem, a more cost-effective strategy is to fine-tune the model with the new interactions in the current time span and inherit the model parameters from the previous time span as the initial values. In this strategy, users' existing intents extracted from previous interactions are not thrown away completely and will be updated in the current time span.
By reconsidering the fundamental concept of fine-tuning an MSR model, we arrive at two intuitions for comprehending the evolution of users' intents over time, encompassing both existing and newly emerging intents, respectively. On one hand, the existing intents of a user, derived from past time periods, remain fluid and may gradually evolve over time. For example, a user initially intents to purchase flip phones might eventually transition their preference to smartphones. On the other hand, the count of newly developed intents can dynamically vary across different time spans. For instance, a user with a prolonged affinity for computer games might suddenly develop an intent in baby care products during a new time period due to the arrival of a child. Building upon these two intuitions, our objective in this paper is to formulate an incremental multi-intent sequential recommendation framework. This framework aims to maintain existing intents through modest adjustment or refinement while adaptively expanding the scope of new intent vectors to accommodate emerging intents. In constructing the aforementioned incremental MSR framework, three challenges remain to be addressed. Initially, existing intents could be forgotten when solely relying on newly collected interactions for model training, leading to performance degradation for items linked to established intents. Through experimentation, we observe that the model's performance diminishes significantly when relying solely on fine-tuning. Secondly, newly conveyed intents from interactions in the current time span might overlap with existing intents due to a fixed model complexity, i.e., a constant number of intent vectors across time spans. Similarly, established intents might reciprocally influence the learning process of new intents. Therefore, it is crucial to prevent any interference between the learning of existing and new intents. Thirdly, distinct users will manifest varying quantities of new intents in each new time span. It is imperative to dynamically determine the appropriate number of new intents for each user within each time span.

The literature for general incremental learning includes reservoir-based methods, regularization-based methods and model-expansion methods.
Reservoir-based methods~\cite{ader,incctr,man} focus on preserving past knowledge by selecting samples from the reservoir for model updating based on prioritizing recency or the extent of being forgotten. However, they need to store historical interactions which may not be obtainable in incremental MSR.
Regularization-based methods~\cite{retrain,asmg} propose to preserve the knowledge learned in the past time spans by enforcing regularization terms, which regularly restrains the model parameters rather than user latent representations.
Moreover, these kinds of methods do not vary model complexity over time, and they cannot generate new intent vectors which require extra model parameters. 
Model-expansion methods~\cite{icarl,pgnn,net2net} aim to expand the model capacity to cope with new knowledge during incremental learning.
While model expansion can be applied to capture new intents and prevent intents from interfering with each other, existing works~\cite{expertgate,net2net} require knowing the expanded model capacity at each time span in advance. Since the number of new intents for each user is dynamically changing, it is a non-trivial task to perform model expansion for incremental multi-intent sequential recommendation. To these ends, existing incremental learning methods are of limited use in tackling the challenges in incremental MSR. 

In this paper, we develop an end-to-end framework for \textbf{I}ncremental \textbf{M}ulti-intent \textbf{A}daptation~(\modelns), which involves the existing-intents retainer, new-intents detector, and projection-based intents trimmer. More specifically, the existing-intent retainer, which aims to solve the first challenge, uses the distillation loss~\cite{kd} to make sure that the existing intent vectors are not far away from their original positions. The new-intents detector is proposed to solve the second challenge, which determines when to create new intents based on the change in distribution of item numbers being classified to different intents. The projection-based intents trimmer is proposed to solve the third challenge, which first allocates a relatively large number of new intents and then modifies the learned intent vectors’ magnitudes and directions to remove redundant intents.
In this way, our framework can preserve the existing intents while an appropriate number of new intents are developed from the new interactions.

However, \modelns, which expands user intent vectors without limitation, may increase the risk of over-fitting and exceed the memory capacity in real recommender systems. On one hand, the intent, which has not been detected in a user's new interactions sequence for a long while, has a lower re-activation probability and can be removed.
On the other hand, some intent vectors located close in space, which represent similar intents, can be compressed as one representative intent vector instead. 
For instance, if a user’s interests include T-shirts, shoes, and sweaters, it would be reasonable to consolidate these clothing-related fine-grained intents into one broader ``clothing" intent, achieving memory efficiency in the process.
Thus, we further upgrade the \model into an \textbf{E}lastic \textbf{M}ulti-intent \textbf{A}daptation version called \modelb which can elastically remove inactive intents and compress user intent vectors under memory space limit.

The main contributions are summarized as follows:
\begin{itemize}[itemsep=0pt,topsep=0pt,parsep=0pt,leftmargin=10pt]
\item We propose a framework for incremental multi-intent sequential recommendation which can dynamically capture new intents from recent interactions for each user while retaining existing intents. 
\item We develop an existing-intents retainer based on knowledge distillation which can preserve existing intents from heavily drifting. We design a new-intents detector to determine when to create new intents and a projection-based intents trimmer to develop new intents adaptively. 
\item We furtherly upgrade the \model into \modelb which can elastically remove inactive intents and compress user intent vectors under memory space limit.
\item We implement our framework on two kinds of multi-intent sequential recommendation models, i.e., dynamic-routing-based and self-attention-based models, and conduct extensive experiments on real datasets. 
The results verify the effectiveness of the proposed \model and \modelb on incremental multi-intent sequential recommendation, compared with various baseline approaches.
\end{itemize}

\section{Background}
Consider a recommendation dataset consisting of interaction records between a set $\mathcal{U}$ of users and a set $\mathcal{I}$ of items. Let $\mathcal{A}=\{(u,i,s)\}$ denote all the interactions in the dataset, where $u\in\mathcal{U}$ is a user id, $i\in \mathcal{I}$ is an item id, and $s$ is the timestamp when the interaction happens. In the sequential recommendation setting, each instance consists of two parts. The input part is $u$'s historical interaction sequence which is a sequence of $u$'s previously interacted items chronically, denoted as $\mathcal{S}_u =\{i_{1,u},i_{2,u},\cdots, i_{n,u}\}$. The output part is the target items $i_{a,u}$ that the user $u$ lastly interacts with. 
Multi-intent sequential recommendation~(MSR) aims to learn a function $f_{\rm user}$ which can map a user's interaction sequence $\mathcal{S}_u$ into a multi-intent user representation ${\boldsymbol H}_u$:
\begin{equation}
\setlength{\abovedisplayskip}{3pt}
\setlength{\belowdisplayskip}{3pt}
{\boldsymbol H}_u = f_{\rm user}(\mathcal{S}_u)=(\boldsymbol{h}_{u,1},\cdots,\boldsymbol{h}_{u,K})\in \mathbb{R}^{d\times K},
\end{equation}
where ${\boldsymbol h}_{u,k}$ denotes the $k^{th}$ intent vector of $u$ ($k\in[1,K]$), $K$ is the number of user intents, and $d$ is the dimension of each intent vector.
The MSR task is to find the top-N candidate items with the highest scores using a scoring function:
\begin{equation}
\setlength{\abovedisplayskip}{3pt}
\setlength{\belowdisplayskip}{3pt}
f_{\rm score}({\boldsymbol H}_u,\boldsymbol{e}_i), 
\end{equation}
where $\boldsymbol{e}_i\in \mathbb{R}^{d\times 1}$ denotes the embedding of item $i\in\mathcal{I}$. 

In practice, user interactions are prolonged with new interactions collected over time spans.
Thus an MSR model should be retrained in each time span to update user representations. Basically, full retraining strategy can be used for MSR, which uses all the historical interaction sequences to retrain the model in each time span. As full retraining is time- and space-consuming when the historical interaction sequences are extremely long, a more cost-effective way is to perform incremental learning with new interactions collected in each time span, which can be formally defined as the incremental multi-intent sequential recommendation. Table~\ref{tab:notation} provides the key notations used throughout the paper.


\begin{definition}[Incremental multi-intent sequential recommendation] 
Let $\mathcal{S}_u^t=\{i_{1,u}^t,i_{2,u}^t,\cdots, i_{n,u}^t\}$ and $i_{a,u}^t$ denote the user $u$'s new interactions and the target item within time span $t$, respectively. In time span $t$, we have an MSR model~$\mathcal{M}^{t-1}$ trained at the previous time span and try to update the model to be $\mathcal{M}^t$ with newly collected sequences $\{\mathcal{S}_u^t|u\in \mathcal{U}\}$.
\end{definition}

To solve the above defined incremental MSR, we have the basic fine-tuning approach based on two kinds of MSR models, namely dynamic-routing based model~(DR) and self-attention-based model~(SA), the detail of which can be found in the conference version~\cite{IMSR}. However, existing intents might be forgotten using the fine-tuning method, which will result in performance degeneration on items related to existing intents. Moreover, new intents may appear in new interactions. If the model capacity is fixed, the occurrence of new intents may interfere with existing intents. Hence, it is desirable to adaptively increase the number of intents to accommodate new intents. In what follows, we describe how the proposed \model resolves these challenges.

\begin{table}[t]
 \renewcommand\arraystretch{0.85}
 \centering
	\caption{Notation table.}
	\label{tab:notation}
	\vspace{-.1in}
	
	\begin{tabular}{ll}
		\toprule
		Notation & Description\\
		\midrule
		$(u,i,t)$ & {the user $u$ interacted with item $i$ at timestamp $t$} \\
		$i_a$ & the target item during training \\
		$\mathcal{U},\mathcal{I}$ & the sets of users and items \\
		$\mathcal{S}_u$ & the historical behavior sequence of user $u$ \\
		$\boldsymbol{h}_{u,k}$ & the $k^{th}$ intent vector for user $u$ \\
		${\bf H}_u$ & the multi-intent representation of user $u$ \\
		$\boldsymbol{e}_a$ & the target item's embedding\\
		\bottomrule
	\end{tabular}

	\vspace{-.23in}
\end{table}

\section{The \model Framework}\label{sec:method}

\subsection{Overview}
\begin{figure}[t]
\vspace{-.1in}
	\centering
	\includegraphics[width=.47\textwidth,height=2.8cm]{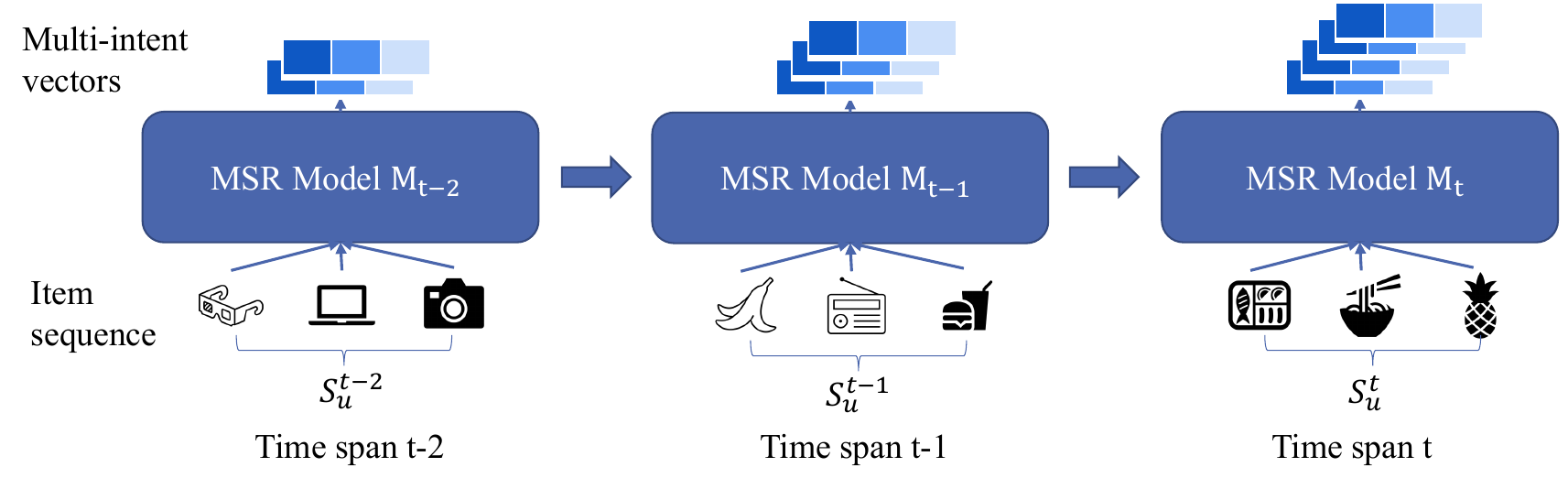}
	\vspace{-.15in}
	\captionsetup{justification=centering}
	\caption{The illustration of incremental learning scheme for multi-intent sequential recommendation.}
	\vspace{-.05in}
	\label{fig:overview}
\end{figure}

The proposed \model is an incremental learning framework for multi-intent sequential recommendation that can dynamically capture new intents from new interactions while retaining existing intents. Figure~\ref{fig:overview} depicts the overall workflow of dynamic multi-intent adaptation. The proposed 
IMA framework contains three modules called existing-intents retainer~(EIR), new-intents detector~(NID), and projection-based new-intents trimmer~(PIT), respectively. EIR is designed to prevent the existing intents learned by the base model in previous time spans from heavily drifting. The new-intents detector creates new intents to prevent the learning of existing and new intents from interfering with each other. The projection-based intents trimmer is proposed to decide the number of new intents adaptively for each user during each time span. 

The remaining parts of this section will be organized as follows. We first introduce the EIR for intent retention in Section~\ref{EIR}. We describe the NID for new-intents detection in Section~\ref{ie} and PIT for the new-intents expansion in Section~\ref{pit}. In Section~\ref{tip} and Section~\ref{imple}, we present the training/inference procedure and the implementation details, respectively.

\subsection{Existing Intents Retention}
\label{EIR}
As stated in the previous works~\cite{mimn,  limarec}, existing and new intents might reappear in the future interactions and we do not know which of the existing or new intents will reappear.
Therefore, it is of great importance to preserve the all existing intents from the previous time span and prevent them from changing greatly or being taken place by new intents.  At the same time, the items belonging to the existing intents can be absorbed into these intents to adjust their representations. The most straightforward idea is to equip a distance-based regularization term like euclidean distance to prevent the existing intent vectors from drifting far away from their original representations after performing multi-intent extractor in a new time span. However, this method is not flexible enough. More specifically, the distance between two vectors of completely different intents can be small, which means even a little change in the euclidean space may change the semantics of an intent vector. 
Thus in worse cases, the distance-based regularization term will prevent items from being recommended by the corresponding intents but cannot prevent the existing intent from changing completely. 

Here we find that knowledge distillation, which does not restrain the representation vector itself but focuses on the restraint of the output logits, is a more flexible way to overcome forgetting problems. Hinton et al. \cite{kd} used knowledge distillation to transfer knowledge from an ensemble of models into a single model for efficient deployment, where knowledge distillation loss is used to preserve knowledge from the cumbersome model by encouraging the outputs of distilled model to approximate that of the cumbersome model. Similarly, the authors of LwF~\cite{LwF} performed knowledge distillation to learn knowledge from new tasks while keeping knowledge on existing tasks in incremental learning scenarios.

Instead of restraining the existing intent representation vector ${\bf h}_k^{(L),t}~(k\in[1,...K_u^{t-1}])$, we propose to use a knowledge distillation loss to overcome the problem. Vanilla distillation loss~\cite{kd} needs a learned teacher model, which is not obtainable in MSR, to guide the student model. However, if we regard the intent vectors as several item classes, the target item preference scoring can also be regarded as a matching-based classification model. Similar to the original knowledge distillation idea, the existing intent vectors can be viewed as the parameters of the teacher model, and the new intent vectors are the parameters of the student model. Thus, a distillation loss to preserve the existing knowledge needs to encourage the outputs of the student model to approximate those of the teacher model. Formally, we have:
\begin{equation}
\setlength{\abovedisplayskip}{3pt}
\setlength{\belowdisplayskip}{3pt}
\begin{aligned}
\label{LKD}
\mathcal{L}_{KD,k,u}^t=\mathcal{L}_{CE}&\left(\sigma\left(\frac{f(\boldsymbol{h}_{k}^{t},\boldsymbol{e}_a^t)}{\tau}\right),\sigma\left(\frac{f(\boldsymbol{h}_{k}^{t-1},\boldsymbol{e}_a^t)}{\tau}\right)\right),\\
\mathcal{L}_{KD}^t &= \sum_{{\mathcal S}_u^t \in \mathcal{D}^t}\sum_{k=1}^{K_u^{t-1}}\mathcal{L}_{KD,k,u}^t,
\end{aligned}
\end{equation}
where $\sigma$ is the sigmoid function, and $\tau$ is the temperature parameter to get soft targets. Here we adopt the dot-product as the matching function $f$. We follow the function of distillation loss in~\cite{incctr}, which replaces the softmax function with the sigmoid function and is efficient to compute. 
In this way, the new intent vectors will keep similar semantics to the existing intents even when their vector is far from the original place in high-level intent vectors' space. We call this component for existing-intents retention as \textit{intents retainer}.

 \subsection{New Intents Detection}
\label{ie}
The previous section proposes a way to preserve the existing intents during incremental learning for MSR. However, a shared and predefined intent number lacks enough diversity and adjustability for the incremental scenario, where the existing intents and new intents will interfere with each other, e.g., the existing intents might be overlaid by new intents or prevent the new intents being captured due to the fixed intents capacity.
Thus, we introduce a new-intents detector in this section which determines when to create new intents based on the distribution of items number being classified to all intents.

 \label{nid}
 \begin{figure}[t]
	\centering
	\subfigure[Skirt]{
		\label{skirt}
		\includegraphics[width=3.9cm,height=2.8cm]{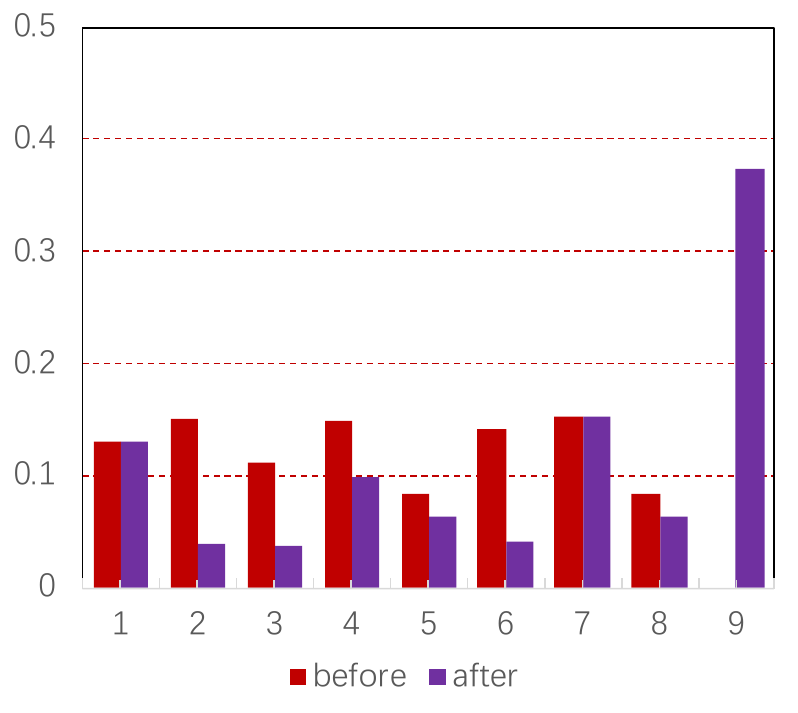}
	}
	\subfigure[LEGO]{
		\label{LEGO}
		\includegraphics[width=3.9cm,height=2.8cm]{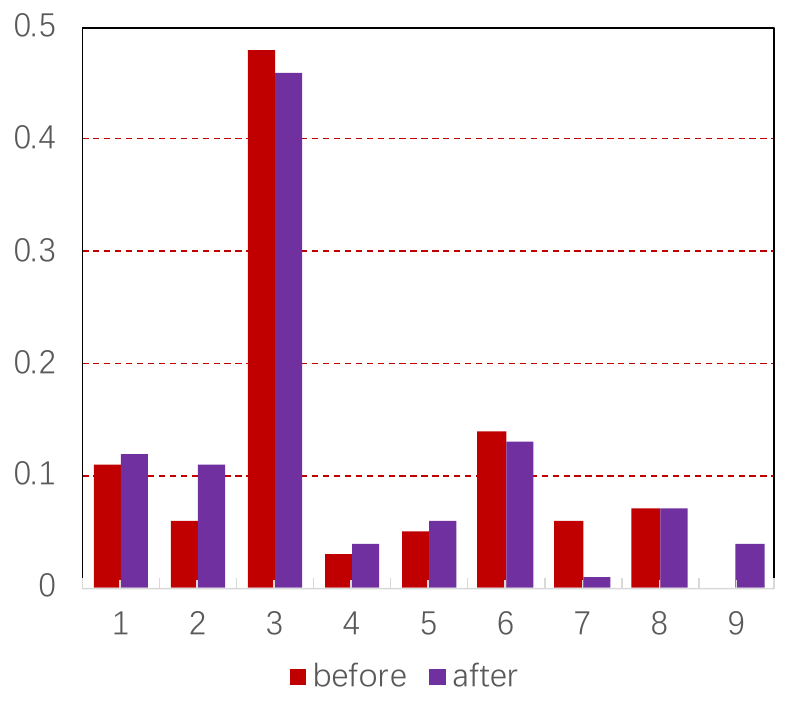}
	}
	\vspace{-.13in}
	\caption{The dot-products of skirt and LEGO to existing/new intents using DR before training~(red) and after training~(purple).}
	\label{fig:puzzle}
	\vspace{-.2in}
\end{figure}

Given a case where a user interacts with both skirt and LEGO in one time span and the user has bought toys but no clothing-related items before, we visualize the dot-products of these two items to the existing intents and new intents. 
In Figure~\ref{fig:puzzle}, we find that skirt has similar dot-products with all eight existing intents and the LEGO has the largest dot-product on the third intent. If we give a new intent vector and retrain the model using fine-tuning strategy, the skirt has the largest dot-product on the newly created ninth intent, and LEGO keeps unchanged. We may infer that the third intent for this user is toy-related, and there are no clothing-related intents. This case illustrated that some items have similar dot-products on all the intents because they are ``puzzled" and can not be classified to any of the intents.

Based on the above observation, we define the \textit{puzzlement} of an item for multi-intent extractor and provide new intent vectors only if most of the items have a certain extent of \textit{puzzlement}.
We use Query Sparsity Measurement~\cite{informer}, which uses KL divergence between the query's attention probability distribution and the uniform distribution to select the most dominant dot-product pairs to represent the \textit{puzzlement}. We do not have a query mechanism in MSR models. However, we can take the dot-product from item embedding $\boldsymbol{e}_i$ to intent vectors $\boldsymbol{h}_k$ in a probabilistic view as:
\begin{equation}
\setlength{\abovedisplayskip}{3pt}
\setlength{\belowdisplayskip}{3pt}
\begin{aligned}
p(\boldsymbol{h}_k|\boldsymbol{e}_i) &= \frac{{\rm kernel} (\boldsymbol{e}_i|\boldsymbol{h}_k)}{\sum_{k'=1}^{K^t}{\rm kernel} (\boldsymbol{e}_i|\boldsymbol{h}_{k'})},\\
{\rm kernel}& (\boldsymbol{e}_i|\boldsymbol{h}_k) = e^{\boldsymbol{e}_i\boldsymbol{h}_k^{\top}},
\end{aligned}
\end{equation}
where $p(\boldsymbol{h}_k|\boldsymbol{e}_i)$ is the posterior probability of the $i^{th}$ item being classified to the $k^{th}$ intent. For brevity, we omit the subscript $u$ and $t$. Intuitively if $p(\boldsymbol{h}_k|\boldsymbol{e}_i)$ is close to a uniform distribution $p(\boldsymbol{h}_k|\boldsymbol{e}_i)=1/K^t$, it means the item can not be well classified into any existing intents and the new intent vectorvectors are needed. Naturally, the distance between two distributions $p$ and $q$ can be used to decide when to create new intents. Here we measure the distance through the Kullback-Leibler divergence as follows:
\begin{equation}
KL(q||p) = {\rm ln} \sum_{k'=1}^{K^t}e^{\boldsymbol{e}_i\boldsymbol{h}_k^{\top}}-\frac{1}{K^t}\sum_{k'=1}^{K^t}\boldsymbol{e}_i\boldsymbol{h}_k^{\top}-{\rm ln} K^t.
\end{equation}

\begin{definition}[Puzzlement]
Given the $i^{th}$ item's embedding and all intent representations, the $i^{th}$ item's \textit{puzzlement} is defined as:
\begin{equation}
P(i) = \frac{1}{K^t}\sum_{k'=1}^{K^t}\boldsymbol{e}_i\boldsymbol{h}_k^{\top}-{\rm ln}\sum_{k'=1}^{K^t}e^{\boldsymbol{e}_i\boldsymbol{h}_k^{\top}}+{\rm ln} K^t,
\label{pi}
\end{equation}
where $K^t$ denotes the intents number in time span $t$, $e_i$ denotes the embedding of the $i^{th}$ item and $h_k$ denotes the representation of the $k^{th}$ intent.
\end{definition}
In the implementation, we will create a predefined number of new intent vectors if the average of all items' \textit{puzzlement} for user $u$ is larger than $c_1$, which is formally defined as:
\begin{equation}
\overline{P(i)}>c_1, i\in \mathcal{S}_u^t.
\label{c1}
\end{equation}
 The set of all these users is the \textit{puzzled} set $\mathcal{U}_{p}^t$. Here the $c_1$ is the hyperparameter that controls the sensitivity of the new-intents detector. We call this component for new-intents detection as \textit{intents detector}.

\subsection{New Intents Expansion}
\label{pit}
\begin{figure}[t]
	\centering
	\subfigure[User1]{
		\label{redundant}
		\includegraphics[width=4.3cm,height=3.2cm]{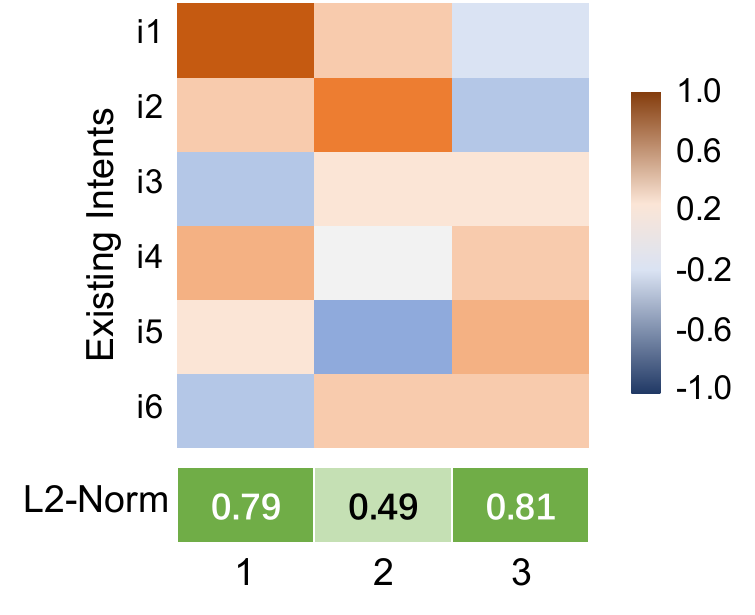}
	}
	\hspace{-.2in}
	\subfigure[User2]{
		\label{nothing}
		\includegraphics[width=4.3cm,height=3.2cm]{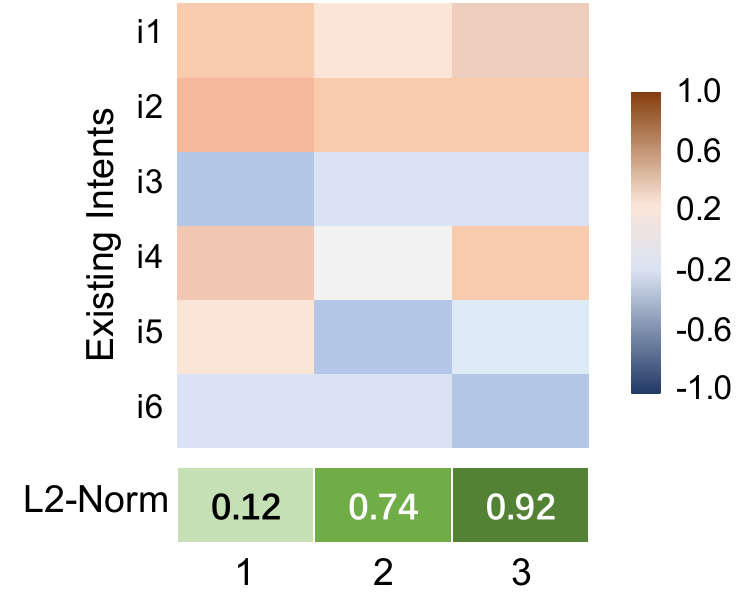}
	}
	\vspace{-.15in}
	\caption{An illustration for two issues of new intents learned without trimming. The upper parts show the Pearson correlation coefficients between existing intents and new intents of two users. The bottom parts show the L2-norm~(Euclidean norm) of new intent vectors.}
	\label{fig:trim}
	\vspace{-.18in}
\end{figure}

The previous section uses \textit{puzzlement} to detect whether the user develops new intents. However, it is not enough because we do not know how many new intents have been developed. Intuitively, there are three ways to decide the number of new intents. The first idea is to create just one intent vector if a new intent is detected each time. This way is easy to implement, but it is not applicable when more than one intent is developed.
An improved idea is to create $\delta K$ new intents certainly. This way is also easy for implementation and more applicable, but it may create too many unnecessary intents in the later time spans and become a great burden to the memory. Moreover, some new intent vectors learn redundant existing intents or even learn nothing. Here are two example users from \textbf{Taobao} that both have 6 existing intents and 3 new learned intents by the model with fixed new-intents expansion number, as shown in Figure~\ref{fig:trim}. 
We compute the dot-product similarities between intent $k$ and $n_u^t$ user's interacted item embeddings and maintain the similarity values for intent $k$ in $\boldsymbol{p}_k=[p_{k,0},p_{k,1},\cdots,p_{k,n_u^t}]$. For any two intents $k$ and $j$, we measure the Pearson correlation coefficient between $\boldsymbol{p}_k$ and $\boldsymbol{p}_j$, which is denoted as $P_{kj}$. L2-Norm denotes the Euclidean norm of each new intent vector.
Higher $P_{kj}$ means higher positive correlation and higher redundant extent for these new intents such as User1's new intent 1 and existing intent 1. And the lower L2-norm means lower existence of learned intents such as User2's new intent 1.

We design a \textit{Projection-based Trimming} mechanism which enables the model to adjustably preserve the right number of new intent vectors for each user which correspond to the new intents developed in the new interaction sequence. The main idea is that we only preserve the orthogonal part of new intent vectors against existing intent vectors' plane and trim the new intent vectors with too small L2-norm. The reasons are two-fold. First, if a new intent vector is close to the existing intent vectors' plane, it intuitively means it represents an intent which is actually the combination of the existing intents, and we do not need to create new vectors for it. Second, the L2-norm of the vector represents the existence of the semantic this intent contains in MSR literature~\cite{MIND,limarec}, which means we can trim the vectors with too small L2-norm because they just learn trivial intents or learn similar intents against the existing ones.

Thus, we first create $\delta K$ new random initialized new intent vectors for each user $u$ at time span $t$. The redundant multi-intent user representation learnt by the extractor becomes:
\begin{equation}
{\boldsymbol H}_{u,rddt}^{t} =(\boldsymbol{h}_{u,1}^{t},\cdots,\boldsymbol{h}_{u,K^t}^{t},\boldsymbol{h}_{u,K^t+1}^{t},\cdots,\boldsymbol{h}_{u,K^t+\delta K}^{t}),
\end{equation}
where ${\boldsymbol H}_{u,rddt}^{t} \in \mathbb{R}^{d\times(K^t+\delta K)}$. $\delta K$ is a hyperparameter shared for all users in all time spans which controls the intent expanding number.
Thus, we employ a projection action onto ${\boldsymbol H}_{u,rddt}^{t}$. Specifically, we project all new intent vectors $\boldsymbol{h}_{u,K^t+1}^{t},\cdots,\boldsymbol{h}_{u,K^t+\delta K}^{t}$ onto the existing intent vectors' plane $span(\boldsymbol{h}_{u,1}^{t},\cdots,\boldsymbol{h}_{u,K^t}^{t})$. The projection can be calculated as:
\begin{equation}
\boldsymbol{h}^{t,proj}_{u,k} = {\boldsymbol M}_{\rm exist}{{\boldsymbol M}_{\rm exist}}^{\top}({\boldsymbol M}_{\rm exist}{\boldsymbol M}_{\rm exist}^{\top})^{-1}\boldsymbol{h}^t_{k},
\label{proj}
\end{equation}
where $ {\boldsymbol M}_{\rm exist}\in \mathbb{R}^{d\times K^t}$ refers to the concatenation of $\boldsymbol{h}_{u,K^t+1}^{t},\cdots,\boldsymbol{h}_{u,K^t+\delta K}^{t}$. In this way, only the component orthogonal to the existing intents will be preserved in the final new intent vector $\boldsymbol{h}^{t}_{k}$ after multi-intent extraction.

For the last step, we will remove the new intent vectors with trivial L2-norm using the following equation:
\begin{equation}
\lVert\boldsymbol{h}^{t}_{k,proj}\rVert_{\rm L2} < c_2.
\end{equation}
In this way, the \textit{Projection-based Trimming} only preserves the orthogonal and new intents with non-trivial existence. The $c_2$ is a hyperparameter that controls the strictness of trivial intent trimming. We call this component for trimming the redundancy intents as \textit{intents trimmer}.
The whole intents expansion algorithm can be found in the conference version~\cite{IMSR}.

\section{\modelnsb: An Elastic Version of \modelns}
\label{sec:ema}
In Section~\ref{sec:method}, we show that the proposed IMA has the capacity to retain existing intents while generating a suitable number of new intents from new interactions. However, \model lacks a constraint on the expansion of user intent vectors, potentially leading to model over-fitting and memory overload in practical recommender systems. On one side, intents that haven't been identified in a user's recent interaction history lose re-activation likelihood and are candidates for elimination. Conversely, closely situated intent vectors, denoting similar intents, can be consolidated into a single representative intent vector. To address these concerns, we enhance \model to \modelb, which dynamically eliminates inactive intents and condenses user intent vectors to fit within memory limitations. In this section, we will first discuss how to find and remove inactive intents and then discuss how to compress similar intent vectors.

\subsection{Inactive Intents Removal}
Intuitively, the intent, which has not been detected in a user's newly collected interactions sequence for a long while, has lower re-activation probability and can be removed. For instance, a user, who has not buying DVDs for a long while, is unlikely to buy DVD in future, because DVD is taken place by stream media. 
Similar to Section~\ref{ie}, we still use inner product $p_{ik}^t$ of item $i$'s embedings and intent $k$'s vectors as the probability of item $i$ being classified to intent $k$ in time span $t$. Intuitively, the intent vector, which has no items being classified to for a long while, can be regarded as inactive intent and be removed. Formally, we define the active score of each intent $k$.
\begin{definition}[Active Score]
Given intent $k$'s vector and all item embeddings, the intent $k$'s \textit{active score} is defined as:
\begin{equation}
AS(k) = \frac{1}{T-t_k}\sum_{t=t_k}^{T}\frac{1}{|\mathcal{S}_t|}\sum_{i\in \mathcal{S}_t}p_{ik}^t,
\label{Ik}
\end{equation}
where $t_k$ denotes the time span when intent $k$ was expanded in, $T$ is the current time span, $S_t$ are the interacted sequence in time span $t$. For brevity, we omit the subscript $u$.
\end{definition}
The intent with lower \textit{active score} is more inactive. For each user, the intents with lowest \textit{active scores} can be removed according to the memory burden. In implementation, we will set a maximal number $K_{max}$ of all intents. We remove the excessive intents with lowest \textit{active scores}.
In implementation, $AS(k)$ can be calculated incrementally. If we stored $tmp_{t-1}=\sum_{t=t_k}^{t-1}\frac{1}{|\mathcal{S}_{t-1}|}\sum_{i\in \mathcal{S}_{t-1}}p_{ik}^t$ in time span $t-1$ for each user, we have $tmp_t=tmp_{t-1}+\frac{1}{|\mathcal{S}_{t}|}\sum_{i\in \mathcal{S}_{t}}p_{ik}^t$ and $AS(k)=\frac{1}{T-t_k}tmp_t$.

\subsection{Similar Intents Compression}
Intuitively, some intent vectors located close in space, which represent similar intents, can be compress as one representative intent vectors instead. For instance, T-shirts, pants and skirts three intents of one user can be compressed into one intent as clothes.
It is straightforward to adopt clustering algorithm to find the similar intents and compress them. Partition-based clustering and density-based clustering are two groups of clustering algorithm. We choose DBSCAN, a density-based clustering algorithm, to implement the similar intents compression in this paper, because we do not want different intents clustered together. 

\begin{algorithm}

\caption{Similar Intents Compression~(SIC)}
\label{alg:compression}
\SetAlgoNoLine
\IncMargin{1em} 
\SetKwInOut{Input}{\textbf{Input}}\SetKwInOut{Output}{\textbf{Output}} 
\Input{
	user intent vectors $\{\boldsymbol{H}_{u}^{t}\}_{u\in\mathcal{U}}$, maximal intent number $K_{max}$\\
	}
\Output{user compressed intent vectors $\{\boldsymbol{H}_{u}^{t,c}\}_{u\in\mathcal{U}}$\\}
\For{\rm{user} $u\in U$}{
	initialize cluster number $k\leftarrow 0$\;
	initialize distance threshold $\epsilon\leftarrow 1.0$\;
	mark all $\boldsymbol{h}_{u}^{t} \in \{\boldsymbol{H}_{u}^{t}\}$ as unvisited\;
	\While{$k<K_{max}$}{
	\For{$\boldsymbol{h}_{u}^{t} \in \{\boldsymbol{H}_{u}^{t}\}$}{
		\If{$\boldsymbol{h}_{u}^{t}$ {\rm is unvisited}}{
		mark $\boldsymbol{h}_{u}^{t}$ as visited\;
		add $\boldsymbol{h}_{u}^{t}$ into cluster $\{\boldsymbol{H}_{u,k}^{t}\}$\;
		add each unvisited vector $\boldsymbol{h} \in \epsilon\text{-}N(\boldsymbol{h}_{u}^{t})\cap\{\boldsymbol{H}_{u}^{t}\}$ to $\{\boldsymbol{H}_{u,k}^{t}\}$, and mark them as visited\;
		$k\leftarrow k+1$\;
		}
	}
	$\epsilon\leftarrow \epsilon/2$\;
	}
	\For{$i \in {\rm range(k)}$}{
		$\boldsymbol{h}_{u,i}^{t}\leftarrow \overline{\{\boldsymbol{H}_{u,i}^{t}\}}$
	}
}
\Return{$\{\boldsymbol{H}_{u}^{t,c}=[\boldsymbol{h}_{u,0}^{t},\cdots,\boldsymbol{h}_{u,k-1}^{t}]\}_{u\in\mathcal{U}}$}\;
\end{algorithm}
Algorithm~\ref{alg:compression} is the procedure of intent compression. $\boldsymbol{h} \in \epsilon\text{-}N(\boldsymbol{h}_{u}^{t})$ means there is a path from $\boldsymbol{h}$ to $\boldsymbol{h}_{u}^{t}$ where each two adjacent nodes have a distance smaller than $\epsilon$. We use the Euclidean Distance as the distance metric. 
$\epsilon$ is automatically set as the value which exactly produces $K_{max}$ intents. Based on our experience, the number $l$ of while loops in Algorithm~\ref{alg:compression} is 3-4.
Because we compress less than $\delta K$ intents in each time span, $\epsilon$ is always a small value which prevents grouping unsimilar intents into one cluster.

\section{Training and Inference Procedure}
\label{tip}
By integrating the \textit{intents retainer}, \textit{intents detector} and \textit{intents trimmer} components together, the whole training process for incremental learning of MSR is as followed. For each user $u$ we initialize $K_u^0$ intent vectors. Then we pretrain a base MSR model on all the user's historical interaction sequences. In this way, we obtain the $K_u^1$ intent vectors for each user, and all the historical data can be discarded. For the time span $2$, we have the new interaction sequence for each user. 
Then we retrain the model for $r$ epochs. 
First, we use the \textit{intents detector} on these new interactions to determine which users have new intents, and we randomly initialize $\delta K$ intent vectors for each of these users.  
Second, we split the user's new interactions into two parts as historical sequences and target items set in time order. For all the users not detected new intents, we compute the user intent vector using multi-intent extractor directly. For users who have new intents detected, we initialize $\delta K$ new intent vectors for them. After this step, we obtain $K_u^2$ intents for each user $u$. 
Third, we calculate the sample softmax loss $\mathcal{L}_{SS}^2$ 
and the knowledge distillation loss $\mathcal{L}_{KD}^2$ using the \textit{intents retainer}. Then we use backpropagation to update the parameters. For users who have new intents detected, we use \textit{intents trimmer} after the multi-intent extractor to preserve the orthogonal part of new intents and trim the new intents with trivial L2-norm. If EMA is used and the intent number exceed the limit $k_{max}$ for a user, we use IIR or SIC to remove or compress redundant intents for this user.
For the next time span, the procedure is identical. In the inference procedure, we calculate the inner-product of $\boldsymbol{v}_u^t$
 and item embedding as the item score. 
Then we evaluate the learned MSR models in the next time span.
The pseudocode for \model and \modelb is provided in Algorithm~\ref{alg1}. The extra parts of \modelb is framed by boxes.

\begin{algorithm}[h]
\SetAlgoNoLine
\IncMargin{1em} 
 initialize 
 base model parameters $\{\boldsymbol{W}^0\}_{base}$\;
 \For{user $u\in U$}{
 	\For{\rm{intent vector} $k\in[1,\cdots,K_u^0]$}{ initialize $\boldsymbol{h}_k^0\leftarrow \mathcal{N}(\boldsymbol{0},\boldsymbol{I})$}
	}
	
	$\{\boldsymbol{W}^t\}_{base}$, $\{\boldsymbol{H}_{u}^{1}\}_{u\in\mathcal{U}}\leftarrow$ pretrain the base model\;
	
	\For{time span $t\in [2, \cdots T]$}{
	$\{\boldsymbol{W}^t\}_{base}$, $\{\boldsymbol{H}_{u}^{t}\}_{u\in\mathcal{U}}\leftarrow \rm{\textbf{Training}}$(
	$\{\boldsymbol{W}^{t-1}\}_{base}$, 
	$\{\boldsymbol{H}_{u}^{t-1}\}_{u\in\mathcal{U}}$)\;
	\tcp{inference}
	 \For{item $i\in\mathcal{S}_u^{t+1}$}{calculate the inner-product of $\boldsymbol{v}_u^t$ and item embedding as the item score;}
	}
 \BlankLine
  \BlankLine
   \BlankLine
 \BlankLine
\BlankLine
\textbf{Training}(
$\{\boldsymbol{W}^{t-1}\}_{base}$, 
$\{\boldsymbol{H}_{u}^{t-1}\}_{u\in\mathcal{U}}, \mathcal{D}_t$):\\

\For{Epoch $i\in [1,\cdots r]$}{
\tcp{ intents expansion}
		$\{\boldsymbol{H}_{u}^{t}\}_{u\in \mathcal{U}}\leftarrow$\\
		\textbf{IntsEx}(
		$\{\boldsymbol{W}^{t-1}\}_{base}$, $\{\boldsymbol{H}_{u}^{t-1}\}_{u\in\mathcal{U}},\mathcal{D}_t$)\;\tcp{intents retention}		  
		 $\mathcal{L}_{SS}^t\leftarrow$get the sample softmax loss\;
		 $\mathcal{L}_{KD}^t\leftarrow$get the distillation loss Eq.(\ref{LKD})\;
		 update 
		  $\{\boldsymbol{W}^{t-1}\}_{base}$ by optimizing $\mathcal{L}_{SS}^t+\mathcal{L}_{KD}^t$\;}
	  
		 $\{\boldsymbol{W}^t\}_{base}\leftarrow$
		 $\{\boldsymbol{W}^{t-1}\}_{base}$\;
		 \tcp{intents removing:}
		 \If{$K>K_{max}$}{
		 
		 \tcp{option 1: inactive intents removal (implemented below)}
		 \fbox{$\{\boldsymbol{H}_{u}^{t,c}\}_{u\in\mathcal{U}}\leftarrow$\textbf{IIR}($\{\boldsymbol{H}_{u}^{t}\}_{u\in\mathcal{U}},k_{max}$)\;}
		 
		 \tcp{option 2: similar intents compression}
		 \fbox{$\{\boldsymbol{H}_{u}^{t,c}\}_{u\in\mathcal{U}}\leftarrow$\textbf{SIC}($\{\boldsymbol{H}_{u}^{t}\}_{u\in\mathcal{U}},k_{max}$)\;}
		 }
		 
		 \Return{
		 $\{\boldsymbol{W}^t\}_{base}$, $\{\boldsymbol{H}_{u}^{t,c}\}_{u\in\mathcal{U}}$}\;
 \BlankLine
 \BlankLine
  \BlankLine
   \BlankLine		 
 \BlankLine	
 \BlankLine
 \fbox{\textbf{IIR}($\{\boldsymbol{H}_{u}^{t}\}_{u\in\mathcal{U}},k_{max}$):}\\
 \For{$u\in\mathcal{U}$}{
 calculate and find $K-K_{max}$ intents $\{\boldsymbol{h}_{u}^{t}\}_x$ with smallest $AS^u(k)$\;
 $\boldsymbol{H}_{u}^{t,c}\leftarrow\boldsymbol{H}_{u}^{t}\backslash\{\boldsymbol{h}_{u}^{t}\}_x$\;
 }
 \Return{$\{\boldsymbol{H}_{u}^{t,c}\}_{u\in\mathcal{U}}$}\;

 \caption{The \modelns/\modelb approach}
 \label{alg1}
\end{algorithm}
\setlength{\textfloatsep}{0.5cm}

\subsection{Implementation Details}
\label{imple}
We implemented our proposed \modelns/\modelb approach using Pytorch 1.8.1 and trained on a 64-bit Linus server equipped with 32 Intel Xeon@2.10GHz CPUs, 128GB memory, and one Titan RTX 2080ti GPU. We choose 64 as the embedding dimension of the item embeddings.  
The batch size is set to 128. 
We choose the Adam optimizer~\cite{adam} to train the model and perform early stopping in the training process. The URL link to our code repository is: https://github.com/Cloudcatcher888/EMA.

\subsection{Complexity Analysis}
The time complexity of EIR, NID, PIT, IIR and SIC is $O(|\mathcal{U}|\overline{K})$, $O(|\mathcal{U}|\overline{|\mathcal{S}_u|})$, $O(|\mathcal{U}|\overline{\delta K})$, $O(|\mathcal{U}|\overline{|\mathcal{S}_u|}\overline{K})$ and $O(|\mathcal{U}|\overline{K})$, which are all much smaller than the time complexity of base model. $|\mathcal{U}|$ and $\overline{|\mathcal{S}_u|}$ denote the user number and average item number of one time span's sequence. $\overline{K}/\overline{\delta K}$ denote the average (extended) intent number. For space complexity, both \model and \modelb require extra space to save $O(|\mathcal{U}|\overline{K})$ intent vectors in former time span.

\section{Experiment}
 \eat{
 To evaluate the performance of our proposed \model framework, we conduct extensive experiments to answer the following research questions:
 \begin{itemize}[itemsep=0pt,topsep=0pt,parsep=0pt,leftmargin=10pt]
 \item \textbf{RQ1} Can \model outperform the existing incremental learning strategies on multi-intent sequential recommendation?
 \item \textbf{RQ2} How does each component of \model contribute to its effectiveness?
 \item \textbf{RQ3} How do the hyperparameters influence the performance of \modelns?
 \item \textbf{RQ4} Where do the improvements of \model come from?
  \item \textbf{RQ5} Can \modelb balance the performance and memory cost?
 \end{itemize}
 }
 In this section, we conduct extensive experiments to evaluate both \model and \modelb framework. Note that we only test \modelb on Xlong, a dataset with extremely long interaction sequences, rather than ordinary dataset used for \modelns. The reason is that there is no need to activate either IIR or SIC for ordinary dataset, on which the result of \model and \modelb will be consequently identical.

\subsection{Experimental Results for \model}
We list the experimental setup of \model including datasets, base models and compared strategies. Then we discuss the overall performance and speed comparison between \model and compared strategies. We also provide the ablation study and parameter sensitivity analysis for \modelns. Note that we omit the case study which can be found in the conference version~\cite{IMSR}.
\subsubsection{Experimental Setup}

\begin{table*}[t]
\centering

 \caption{Statistics of the datasets.}
 \vspace{-.1in}
 \label{tab:dataset}
 \begin{tabular}{c|c|c|ccccccc}
\toprule
 \multirow{2}{*}{Dataset} & \multirow{2}{*}{\#users} & \multirow{2}{*}{\#items} &\multicolumn{7}{c}{\#interactions}\\
 & & & pre-training&1&2&3&4&5&6 \\ 
 \midrule
 Electronics & 87,912 & 234,621 & 1,689,188 & 224,421&428,149 &329,194 &129,482 &481,491 & 196,451\\
 Clothing & 285,464 & 376,859 & 5,748,920 & 864,371&574,922 &957,329 &1,134,792 &943,422 & 1,274,084\\
 Books & 459,133 & 313,966 & 8,898,041 & 1,345,234& 1,324,545& 1,852,324& 1,593,281& 1,349,281& 1,433,376\\
 Taobao & 976,779 & 1,708,530 & 85,384,110 & 12,329,481& 14,481,123& 22,129,123& 9,329,128& 14,238,129& 12,877,126\\
\bottomrule
 \end{tabular}
 \vspace{-.2in}
\end{table*}

\noindent\textbf{Datasets} We use four real-world sequential recommendation datasets from Amazon\footnote{http://jmcauley.ucsd.edu/data/amazon/} and Taobao\footnote{https://tianchi.aliyun.com/dataset/dataDetail?dataId=649\&userId=1}, which are adopted by many existing MSR works~\cite{MIND,ComiRec}.
\begin{itemize}[itemsep=0pt,topsep=0pt,parsep=0pt,leftmargin=10pt]
 \item \textbf{Amazon}: This dataset consists of reviews of different kinds of products from Amazon~\cite{amazon1,amazon2}. We consider three categories of products and obtain \textbf{Electronics}, \textbf{ Clothing} and \textbf{Books} subsets. We use the item id and the UNIX review time from the metadata. 
 \item \textbf{Taobao}: The dataset is collected from the e-commerce platform Taobao~\cite{taobao}. In our experiment, we only use the click behaviors and sort one user's behaviors by time. 
 \end{itemize}
 For a fair comparison, we follow the same data preprocessing and data splitting rule for all the approaches. 
We discard all users with fewer than 30 interactions.
 The whole timeline $[0,Z]$ is split into $T+1$ time spans, where the first time span is $[0,\alpha Z]$ and $[\alpha Z,Z]$ is equally divided into $T$ time spans. We set $T=6$ and $\alpha=0.5$ for all the datasets. Each user interaction is assigned to the corresponding time span. 
 The interactions whose timestamps are in the range of $[0, \alpha Z]$ compose the pre-training dataset, while the interactions in the following $T$ time spans are $T$ incremental datasets. 
 For each user in each time span, we use the latest interaction for testing, the second last interaction for validation, and all the remaining interactions for training. 
 For the full retaining strategy, we use the pre-training dataset and $1^{th},2^{th},...,t^{th}$ incremental datasets to retrain model in the time span $t$. For the incremental learning strategies~(including \modelns), we use the $t^{th}$ incremental dataset to fine-tune model in the time span $t$. We test model on the $(t+1)^{th}$ incremental dataset. We have excluded the test performance of the pretrained model (in $0^{th}$ time span) and computed the average performance over $1^{th},2^{th},...,{(T-1)}^{th}$ time spans.
 The statistics of the datasets are listed in Table~\ref{tab:dataset}.
 \label{datasets}

\noindent\textbf{Evaluation}
We use the hit ratio~(HR), and NDCG on Top20 as the metrics to evaluate the performance of all the comparison methods, which are commonly used in multi-intent sequential recommendation~\cite{psr,bert4rec}.

\noindent\textbf{Base Models} We consider two dynamic-routing-based MSR models and one self-attention-based MSR model, which are listed as follows.
\begin{itemize}[itemsep=0pt,topsep=0pt,parsep=0pt,leftmargin=10pt]
\item \textbf{MIND}~\cite{MIND} is a typical dynamic-routing based model for multi-intent sequential recommendation. It uses a simplified transformation matrix called shared bilinear mapping matrix to extract multiple intent representations from user interaction sequences. The routing logits are initialized randomly. The default number of intent representations $K_u^t$ is set to 4.
\item \textbf{ComiRec-DR}~\cite{ComiRec} is another dynamic-routing based model for MSR. It uses a shared affine transformation matrix to extract intent representations from user interaction sequences. The routing logits are initialized as zero. The default number of intent representations $K_u^t$ is set to 4.
\item \textbf{ComiRec-SA}~\cite{ComiRec} is a self-attention based model for MSR. It employs a multi-head self-attention module to extract intent representations from user interaction sequences, where each head represents one intent. The default number of heads $K_u^t$ is set to 4.
\end{itemize}
We choose 64 as the dimension of intent representations for all three models on all the datasets.

\noindent\textbf{Compared Learning Strategies} We compare our proposed \model with the full retraining strategy and three existing incremental learning strategies on the above three base models. 
 \begin{itemize}[itemsep=0pt,topsep=0pt,parsep=0pt,leftmargin=10pt]
 \item \textbf{Full retraining~(FR)} In each time span, all the model parameters will be reinitialized and the whole user historical interaction sequences will be used to retrain the base model. The intents number will be kept same as \model.
  \item \textbf{Fine-tuning~(FT)} In each time span, all the parameters will be inherited from the previous time span and only the newly collected user interactions will be used to fine-tune the base model.
 \item \textbf{SML}~\cite{sml} is a model-based incremental learning approach that employs a CNN-based transfer module to leverage the previous model knowledge while training the current model on $\mathcal{D}_t$. We use $5\times5$ filters. The best results are reported by choosing the number of CNN filters in $\{2, 5, 8, 10\}$ and the MLP hidden size in $\{10, 20, 40, 80\}$. 
 \item \textbf{ADER}~\cite{ader} is a sample-based incremental learning approach for the session-based recommendation that samples historical interaction sessions from a session pool according to cosine similarity with the newly collected sessions as the complementary set for new sessions, which is useful to preserve the user's long-term intents. We add 5 randomly truncated interaction sequences to the session pool for each user in each time span.
 \end{itemize}
 	For a fair comparison, all compared learning methods are optimized with the loss function.
    The regularization coefficient for $\mathcal{L}_{KD}$ is tuned in $\{1e{-2},1e{-3},\cdots,1e{-6},0\}$. 
	The learning rate is tuned in $\{0.1,0.01,0.005,0,001\}$. 
	The incremental training epoch is tuned in $\{5,10,15,20,30,50\}$. 

\subsubsection{Overall Results}

\begin{table*}[t]
\centering
\vspace{-.0in}

 \caption{ The performance comparison results. * denotes $p<0.05$ when performing the two-tailed pairwise t-test on \model with the incremental learning methods~(SML or ADER). The \textbf{bold} and the \underline{underline} show the best and second-best results within four incremental learning methods, respectively. RI indicates a relative improvement of the average score of HR and NDCG against FT. All the numbers in the table are percentage numbers with `\%’ omitted.
 }

 	\vspace{-.1in}
	\setlength{\tabcolsep}{1.95mm}{
	
 \begin{tabular}{cc|cccccccccccc}
\toprule
  \multirow{2}{*}{Base model}\!& \multirow{2}{*}{Learning method} &\multicolumn{3}{c}{Electionics}&\multicolumn{3}{c}{Clothings}&\multicolumn{3}{c}{Books}&\multicolumn{3}{c}{Taobao}\\
 & & HR & NDCG & RI & HR & NDCG & RI & HR & NDCG & RI & HR & NDCG & RI \\ 
 \midrule										
\multirow{5}{*}{MIND} 																								
& FR				&16.03 			&16.43 			&11.15 			&16.23 			&15.98 			&10.57 			&13.82	 		&11.95 			&10.47 			&43.29 			&24.90 			&2.63 	\\
& FT				&14.75 			&14.46 			&-				&14.45 			&14.68 			&-				&12.34 			&10.98 			&-				&42.09 			&24.35 			&-		\\
& SML			&15.41 			&\underline{15.17} 	&4.71 			&15.27 			&14.81 			&3.26 			&\underline{13.12} 	&11.12 			&3.97 			&42.88 			&\underline{24.58} 	&\underline{1.54} 	\\
& ADER			&\underline{15.64} 	&14.98 			&\underline{4.84} 	&\underline{15.62} 	&\underline{15.20} 	&\underline{5.76} 	&12.92 			&\underline{11.48} 	&\underline{4.64} 	&\underline{42.90} 	&24.24 			&1.05 	\\
& \textbf{\model}		&15.81* 			&15.71* 			&\textbf{7.93} 		&15.81* 			&15.71* 			&\textbf{8.19}			&13.99* 			&11.94* 			&\textbf{11.18} 			&43.94* 			&25.66* 			&\textbf{4.76} 	\\
\midrule	\multirow{5}{*}{ComiRec-DR} 																							
& FR				&17.00 			&16.79 			&9.85 			&16.91 			&16.75	 		&9.82 			&14.79	 		&12.79 			&12.06 			&44.29 			&25.87 			&4.23 	\\
& FT				&15.41 			&15.35 			&-				&15.36 			&15.28 			&-				&13.30 			&11.30 			&-				&42.62 			&24.68 			&-		\\
& SML			&\underline{16.16} 	&15.85 			&4.09 			&\underline{16.08} 	&15.77 			&3.92 			&\underline{13.92} 	&11.85 			&\underline{4.74} 	&43.28 			&24.89 			&1.28 	\\
& ADER			&16.12 			&\underline{15.90} 	&\underline{4.10} 	&16.02 			&\underline{15.84} 	&\underline{3.96} 	&13.73 			&\underline{11.96} 	&4.43 			&\underline{43.44} 	&\underline{25.00} 	&\underline{1.68} 	\\
& \textbf{\model}		&16.80* 			&16.48* 			&\textbf{8.20} 			&16.74* 			&16.47* 			&\textbf{8.38} 			&14.46* 			&12.48*	 		&\textbf{9.51} 			&44.48* 			&26.00* 			&\textbf{4.72} 	\\
\midrule	\multirow{5}{*}{ComiRec-SA} 																									
& FR				&17.15 			&16.95 			&10.82 			&16.74	 		&16.87	 		&8.83 			&14.86 			&12.85	 		&11.66	 		&44.31 			&25.75 			&4.54 	\\
& FT				&15.31 			&15.46 			&-				&15.49 			&15.39 			&-				&13.46 			&11.35 			&-				&42.44 			&24.58 			&-		\\
& SML			&15.96 			&\underline{15.99} 	&3.83 			&15.90 			&15.88	 		&2.89 			&\underline{13.78} 	&11.71 			&2.72 			&43.17 			&24.83 			&1.47 	\\
& ADER			&\underline{16.32} 	&15.88 			&\underline{4.63} 	&\underline{16.14} 	&\underline{15.88} 	&\underline{3.67} 	&13.55 			&\underline{11.98} 	&\underline{2.87} 	&\underline{43.43} 	&\underline{25.00} 	&\underline{2.12} 	\\
& \textbf{\model}		&16.97* 			&16.32* 			&\textbf{8.19} 			&16.94* 			&16.56* 			&\textbf{8.45} 			&14.38* 			&12.49* 			&\textbf{8.30} 			&44.58* 			&26.11* 			&\textbf{5.48} 	\\
\bottomrule
 \end{tabular}
 
 }
 \label{tab:result}
\vspace{-.18in}

\end{table*}

\begin{table}
\centering

\caption{ The performance comparison results between \model and life-long MSR models. The average HR over 5 time spans are reported.}
 \label{tab:msrmodel}
 \setlength{\tabcolsep}{1.5mm}{
\begin{tabular}{c|cccc}
\toprule
Datasets&Electronics&Clothings&Books&Taobao\\
\midrule
MIMN&14.11&14.37&11.87&41.02\\
LimaRec&15.31&15.02&13.07&42.33\\
\modelns~(ComiRec-DR)&16.81&16.68&14.48&44.35\\
\bottomrule
\end{tabular}
}
\vspace{-.15in}
\end{table}

\begin{figure*}[t]
	\centering
\begin{minipage}[b]{\linewidth}
	\begin{subfigure}
   \centering
	\includegraphics[width=.95\textwidth]{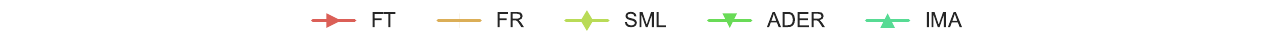}
	\end{subfigure}
	\vspace{-.11in}
	\end{minipage}
	\setcounter{subfigure}{0}
	\begin{minipage}[b]{\linewidth}
	\subfigure[Electronics-HR]{
			\includegraphics[width=4.3cm]{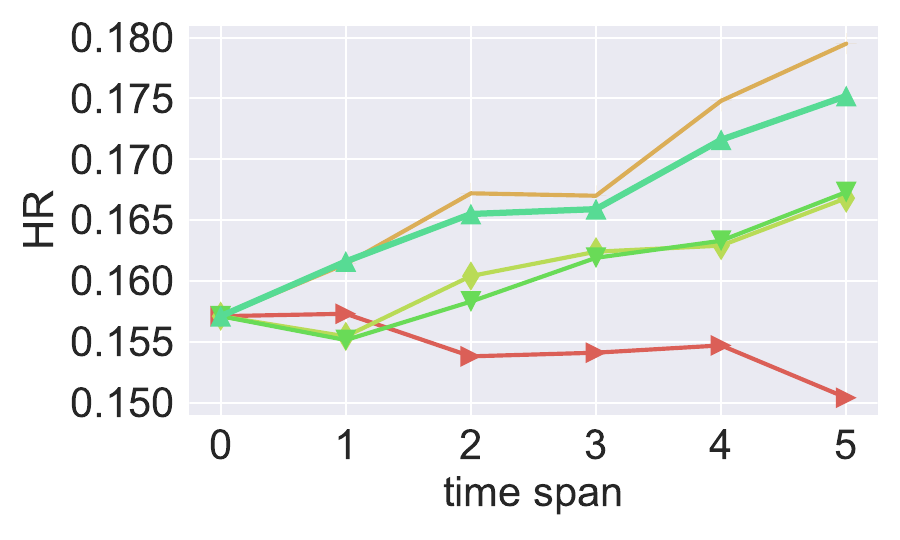}
	}
	\hspace{-.1in}
	\subfigure[Clothing-HR]{
			\includegraphics[width=4.3cm]{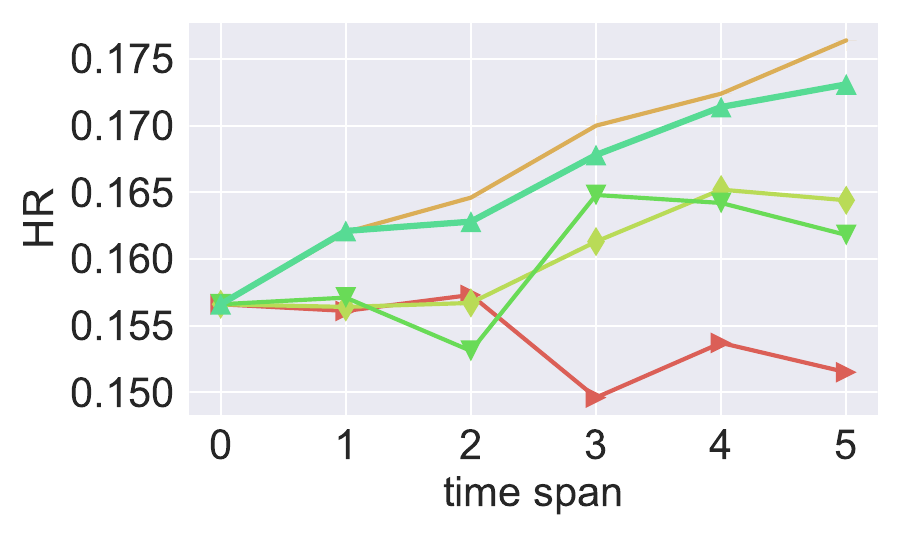}
	}
	\hspace{-.1in}
	\subfigure[Books-HR]{
			\includegraphics[width=4.3cm]{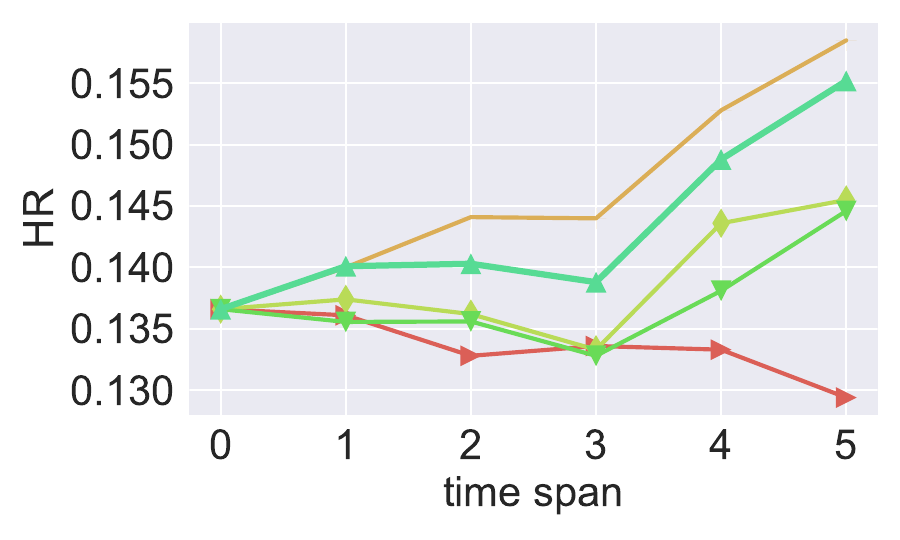}
	}
	\hspace{-.1in}
	\subfigure[Taobao-HR]{
			\includegraphics[width=4.3cm]{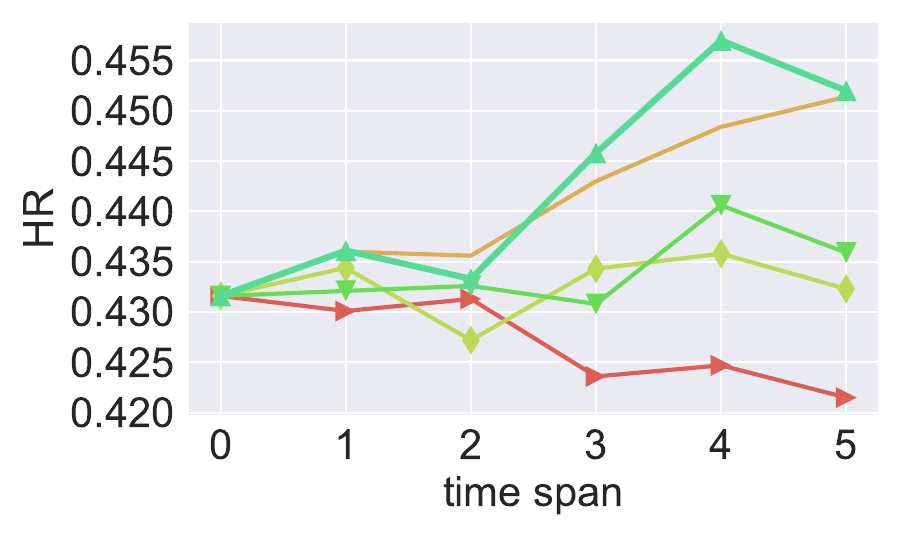}
	}
	\vspace{-.05in}
	\end{minipage}
	\begin{minipage}[b]{\linewidth}
	\subfigure[Electronics-NDCG]{
			\includegraphics[width=4.3cm]{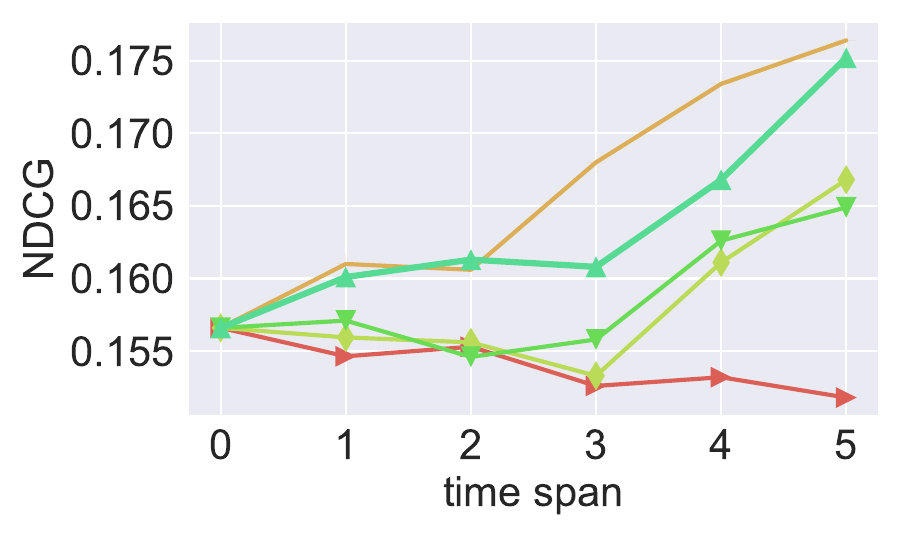}
	}
	\hspace{-.1in}
	\subfigure[Clothing-NDCG]{
			\includegraphics[width=4.3cm]{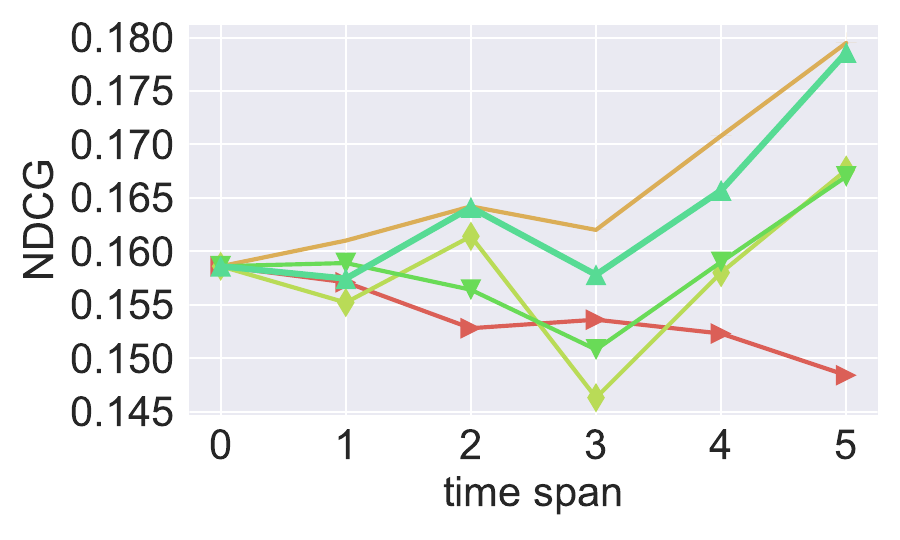}
	}
	\hspace{-.1in}
	\subfigure[Books-NDCG]{
			\includegraphics[width=4.3cm]{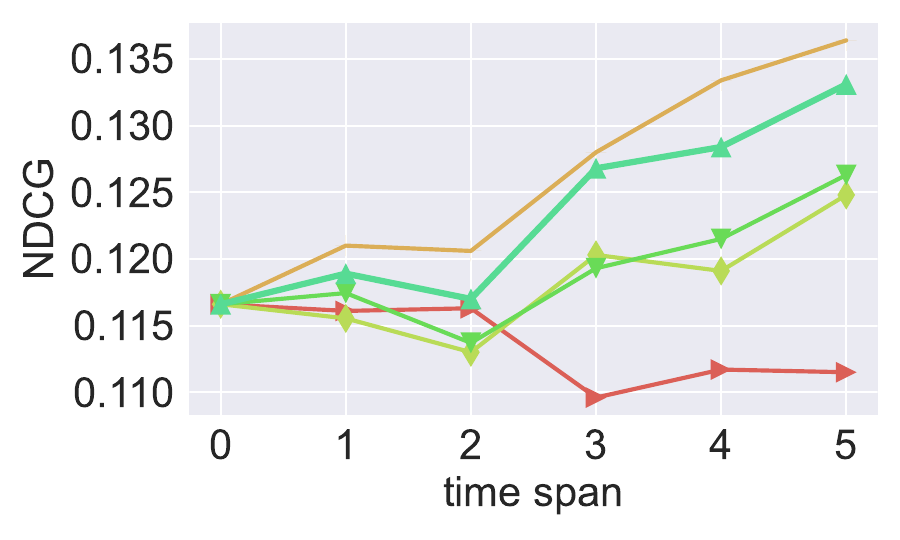}
	}
	\hspace{-.1in}
	\subfigure[Taobao-NDCG]{
			\includegraphics[width=4.3cm]{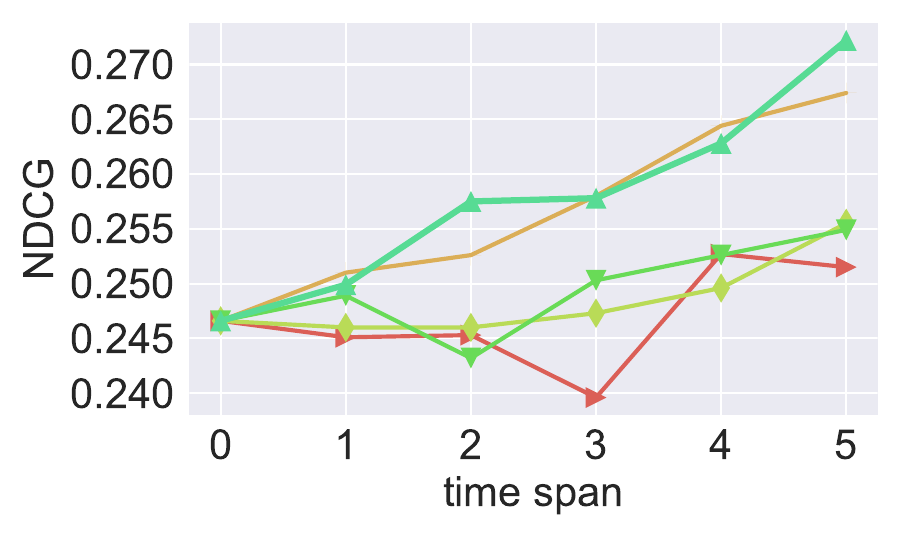}
	}
	\vspace{-.2in}
	\end{minipage}
	\caption{HR and NDCG on each time span. }
	\vspace{-.15in}
	\label{fig:result}
\end{figure*}

\noindent\textbf{Performance Comparison}
Table~\ref{tab:result} provides the performance comparison results of different methods on four datasets.Each result is averaged by 10 repeated experiments. 

We have the following observations.
Table~\ref{tab:result} shows SML and ADER outperform FT in most scenarios. The reason is that they both partially preserve the existing intents by sampling truncated historical interaction sequences~(ADER) or transferring knowledge from previous model parameters~(SML). However, their performances are inferior to FR, which leverages the whole user interaction sequences.
Table~\ref{tab:result} shows \model achieves
3.77\%,
3.89\%,
4.21\%,
4.76\%
relative improvements on NDCG compared to the second best incremental learning methods on four datasets~(averaged on three base models), respectively. The reason for the performance improvement is that \model can capture new intents with NID and PIT while SML and ADER do not expand the model capacity over time. Moreover, the performance improvements on different base models are close, which reflects that \model is effective in performing incremental multi-intent recommendation on different kinds of MSR models. 
 Figure~\ref{fig:result} gives the detailed performance trends of different methods over time spans using ComiRec-DR~(similar trends can be found on other base models). 
The performance of FT decreases significantly over time spans. The results of SML and ADER also drop fast, which also testifies that SML and ADER cannot losslessly preserve existing intents from historical interactions. Moreover, the result of \model drops slightly faster than FR by only using the newly collected interactions thanks to the new-intents expansion mechanism, which alleviates the existing intents forgetting problem.
Figure~\ref{fig:result} shows that the performances of all the compared incremental learning methods become worse on \textbf{Taobao} except \modelns. The reason is that \textbf{Taobao} has more items and users' intents change more rapidly, which will amplify the superiority of \model with the capability of new-intents expansion.  
Table~\ref{tab:msrmodel} provides the HR results between \model based on ComiRec-DR and life-long MSR models~(includes MIMN~\cite{mimn} and LimaRec~\cite{limarec}). MIMN adopts Neural Turing Machine to adaptively read or write user intents according to the online interactions. LimaRec employs linear self-attention to identify relevant information from users’ interaction sequences with different intents. These two works differ from ours in the sense that they focus on incrementally updating users’ representations during online inference and do not provide an incremental learning method for MSR model updating. On average, \model achieves 
3.61\%, 
2.89\%,
5.12\%, 
4.47\% 
relative improvements on HR compared to the best life-long MSR model~(LimaRec) on the four datasets, respectively. We observe that the life-long MSR models perform inferior to \model because they only update user representations but do not update the model parameters after pretraining. Another reason for the performance gap is that life-long MSR models use fixed number of intents while \model can adaptively create vectors to capture newly evolved intents.
 
 \noindent\textbf{Speed-up} We compare the training time of different methods at different time spans and average inference time over all time spans. Table~\ref{tab:speedup} shows the time cost on \textbf{Taobao}. Similar conclusions can be drawn from the other datasets. 
The training time of FR~(on MIND and ComiRec-DR) and ADER grows linearly due to the increasing lengths of interaction sequences or the larger pool size of truncated historical interaction sequences. The training time of FR on ComiRec-SA increases more rapidly because the self-attention module has quadratic time complexity. The training time of SML is stable across different time spans because it does not vary the model capacity. However, SML requires longer training time in each time span due to the high computational complexity of its meta-learner.
\model is about 6 times faster than FR~(on MIND and ComiRec-DR), and the retraining time is stable across different time spans. The \model uses 3.5\% extra training time compared with FT to achieve significant performance improvement.  We find the inference time depends on both the base model and the number of intents, where \model takes slightly longer inference time~(at the 100ms level per instance) due to the adaptive number of intents.
 \vspace{-.06in}

\begin{table}[t]
\centering

 \caption{Training/Inference time (in seconds) on \textbf{Taobao} dataset.}
 \vspace{-0.1in}
 \label{tab:speedup}
\setlength{\tabcolsep}{.7mm}{
 \begin{tabular}{cc|ccccc|c}
\toprule
 \multirow{2}{*}{\shortstack{Base\\ Model}}&\multirow{2}{*}{\shortstack{Learning\\ Method}}&\multirow{2}{*}{t=1}&\multirow{2}{*}{t=2}&\multirow{2}{*}{t=3}&\multirow{2}{*}{t=4}&\multirow{2}{*}{t=5} &  \multirow{2}{*}{\shortstack{Average\\ Inference Time}}\\ 
 & & & & & & & \\
 \midrule
\multirow{5}{*}{MIND} &FR & 4,671 & 4,877 & 4,891 & 5,189 & 5,382 &\multirow{4}{*}{0.15}\\
& FT & 738 & 819 & 802& 822 & 799& \\
 & SML & 908 & 922 & 941 & 913 & 902 &\\
 &ADER & 982 & 1,154 & 1,379 & 1,534 & 1,712 &\\
 \cmidrule{8-8}
& \model & 811 & 832 & 851 & 804 & 823 &0.17\\
  \midrule
\multirow{5}{*}{ComiRec-DR} & FR & 5,472 & 5,693 & 5,871 & 5,902 & 6,023 &\multirow{4}{*}{0.17}\\
 &FT & 928 & 949 & 932& 941 & 946& \\
 & SML & 1,052 & 1,098 & 1,079 & 1,073 & 1,081& \\
 &ADER & 990 & 1,199 & 1,499 & 1,591 & 1,891 &\\
  \cmidrule{8-8}
 & \model & 941 & 962 & 954 & 994 & 983&0.21\\
  \midrule
\multirow{5}{*}{ComiRec-SA} & FR & 5,569 & 6,214 & 7,112 & 8,219 & 9,401&\multirow{4}{*}{0.31}\\
 &FT & 972 & 991 & 992& 1,001 & 982 &\\
 & SML & 1,111 & 1,121 & 1,141 & 1,101 & 1,102 &\\
 &ADER & 1,054 & 1,231 & 1,529 & 1,681 & 1,952 &\\
  \cmidrule{8-8}
 & \model & 1,012 & 1,031 & 1,041 & 1,104 & 1,083 &0.33\\

\bottomrule
 \end{tabular}
 }
\vspace{-.05in}
\end{table}


\subsubsection{Ablation Study}

\begin{figure}[t]
	\centering
	\begin{minipage}[b]{\linewidth}
		\begin{subfigure}
   \centering
	\includegraphics[width=\textwidth]{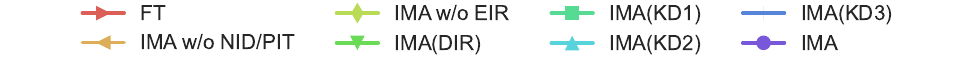}
	\end{subfigure}
	\vspace{-0.15in}
	\end{minipage}
	\setcounter{subfigure}{0}
	\begin{minipage}[b]{\linewidth}
	\centering
	\subfigure[Books-ComiRec-DR]{
			\includegraphics[width=4.2cm]{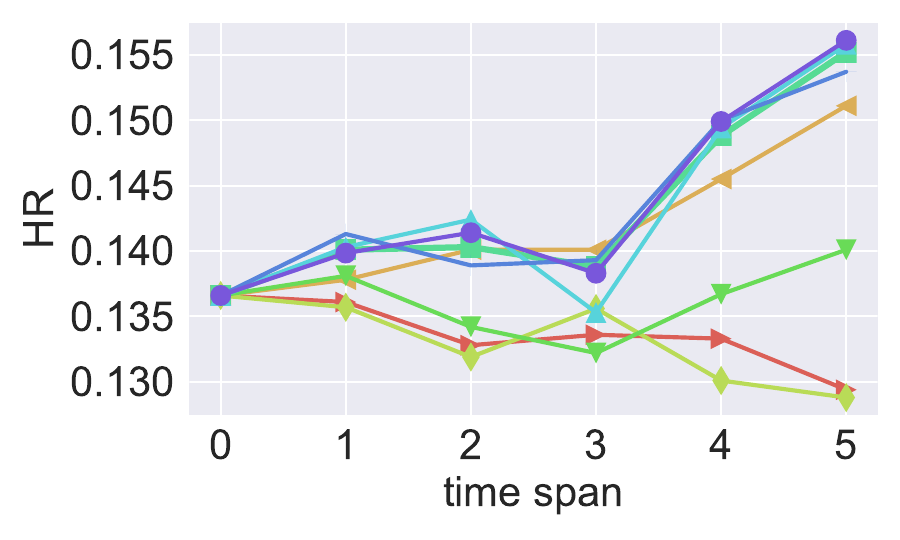}
	}
	\hspace{-.15in}
	\subfigure[Books-ComiRec-SA]{
			\includegraphics[width=4.2cm]{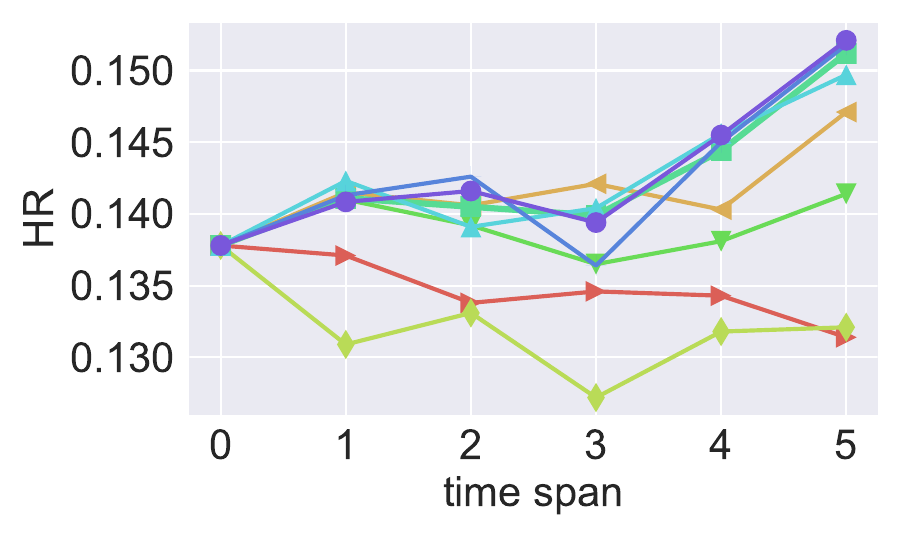}
	}
	\vspace{-0.05in}
	\end{minipage}
	\begin{minipage}[b]{\linewidth}
	
   \centering
	\setcounter{subfigure}{2}
	\subfigure[Taobao-ComiRec-DR]{
			\includegraphics[width=4.2cm]{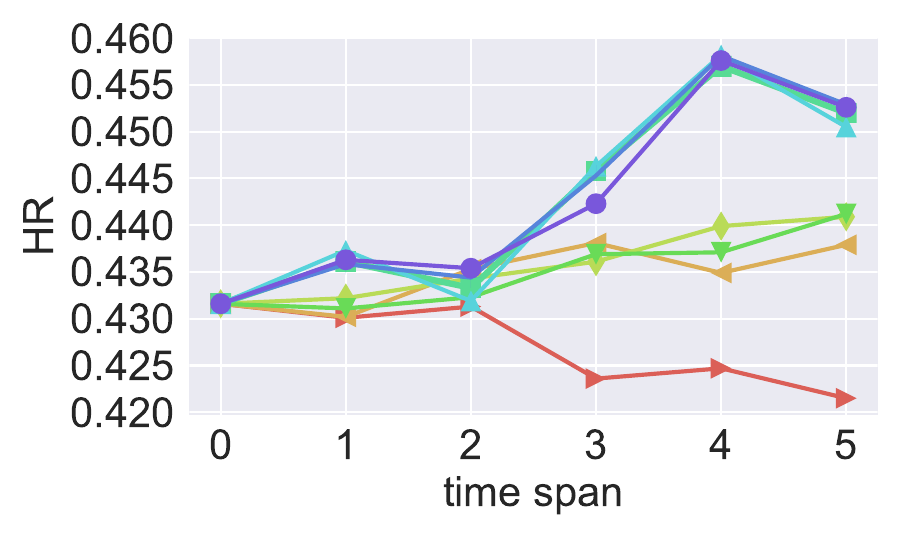}
	}
	\hspace{-.15in}
	\subfigure[Taobao-ComiRec-SA]{
			\includegraphics[width=4.2cm]{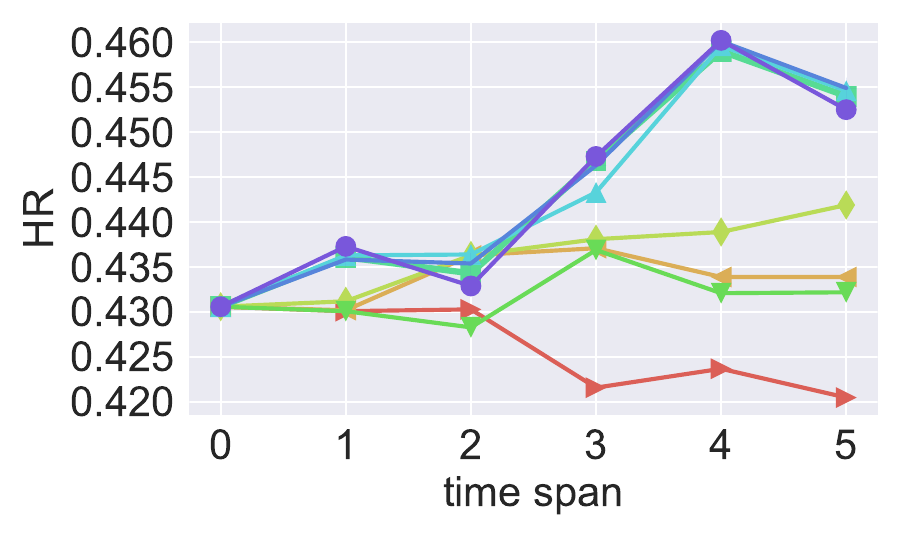}
	}
	\end{minipage}
\vspace{-.15in}
	\caption{Ablation Study on \textbf{Books} and \textbf{Taobao}.}
	\label{fig:ablation}
	\vspace{-.15in}
\end{figure}

We perform an ablation study on the two largest datasets, \textbf{ Books} and \textbf{Taobao}, using ComiRec-DR/SA to evaluate the effects of different components in \model on the recommendation performance. 
Specifically, we consider the following variants.
\begin{itemize}[itemsep=0pt,topsep=0pt,parsep=0pt,leftmargin=10pt]
 \item \textbf{FT:} Using the base model with FT as the training method.
 \item \textbf{\model w/o {NID}\&{PIT}:} Removing {NID} and {PIT} from \modelns.
 \item \textbf{\model w/o {EIR}:} Removing the {EIR} module from \modelns.
 \item \textbf{\modelns({DIR}):} Replacing the {EIR} with the {DIR} for intent retention, which changes the Eq.~(\ref{LKD}) with the Euclidean distance-based regularization term~\cite{kd}.
 \item \textbf{\modelns(KD1/KD2/KD3):} Replacing Eq.~(\ref{LKD}) with three softmax-based distillation losses~\cite{sakd,LwF,lsil} 
 to evaluate the effects of different distillation losses on the recommendation performance.

\end{itemize}
As shown in Figure~\ref{fig:ablation}, 
 \model performs best among all the comparison methods both on DR and SA models, which shows that the removal of any component from \model will hurt the final performance, and the contribution of all three components is insensitive to the base model. 
 On \textbf{Taobao}, the effectiveness of the {NID} and {PIT} is significant, which reflects that users in \textbf{Taobao} develop new intents fast due to the richness of item categories. Specifically, we observe the average intents number of all users in \textbf{Taobao} increases from 4.0 to 9.2 with \modelns. 
However, on \textbf{Books}, the effectiveness of the {EIR} is more significant, which reflects that users' intents in books are more stable, and it is more important to preserve users' existing intents. As evidence, the average intents number of all users in \textbf{Books} only increases from 4.0 to 5.6 with \modelns.
On \textbf{Books}, the removal of {EIR} makes the performance of \model even inferior to FT. The performance decrease is also significant on \textbf{Taobao}. This reflects that preserving existing intents is the foundation for new-intents detection
. We can see that using a distance-based regularization term for intent retention has the inferior performance to using knowledge-distillation-based regularization. We also find \modelns(DIR) performs worse on items which do not exactly match with existing intents. For instance, \modelns(DIR) fails to recommend smartphones for a user who had intent to flip phones, while \model works. The above two observations show that {EIR} is more flexible than {DIR} on existing intents preserving. Nevertheless, different kinds of knowledge distillation terms achieve similar performance, which testifies that our method is insensitive to the choice of the distillation loss function.

\begin{figure*}[t]
\setlength{\abovecaptionskip}{0.1cm}
	\centering
	\begin{minipage}[b]{2\linewidth}
		\begin{subfigure}
   \centering
	\includegraphics[width=.5\textwidth]{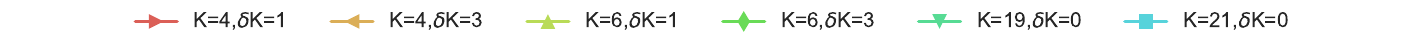}
	\end{subfigure}
	\vspace{-0.15in}
	\end{minipage}
	\setcounter{subfigure}{0}
	\begin{minipage}[b]{2\linewidth}
	\subfigure[Books-DR-c1]{
			\includegraphics[width=4.3cm,height=2.6cm]{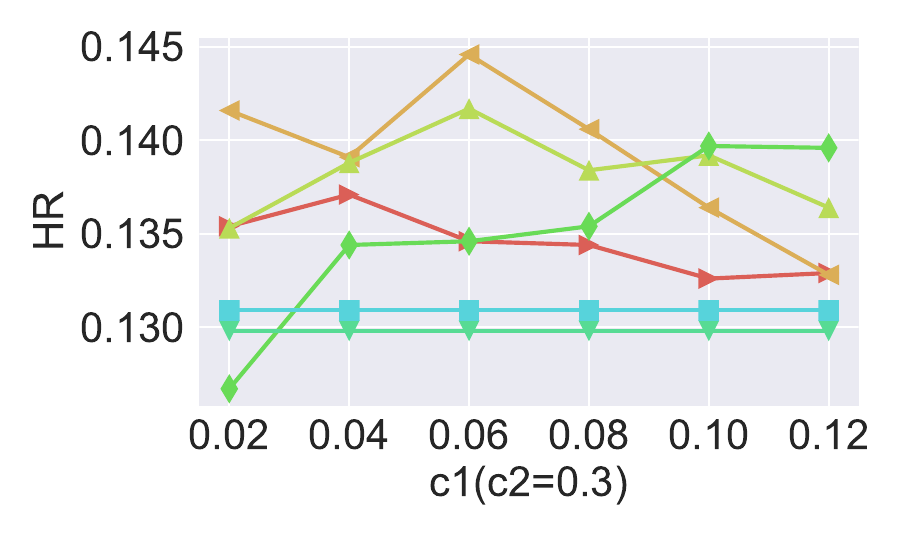}
	}
	\hspace{-.1in}
	\subfigure[Books-SA-c1]{
			\includegraphics[width=4.3cm,height=2.6cm]{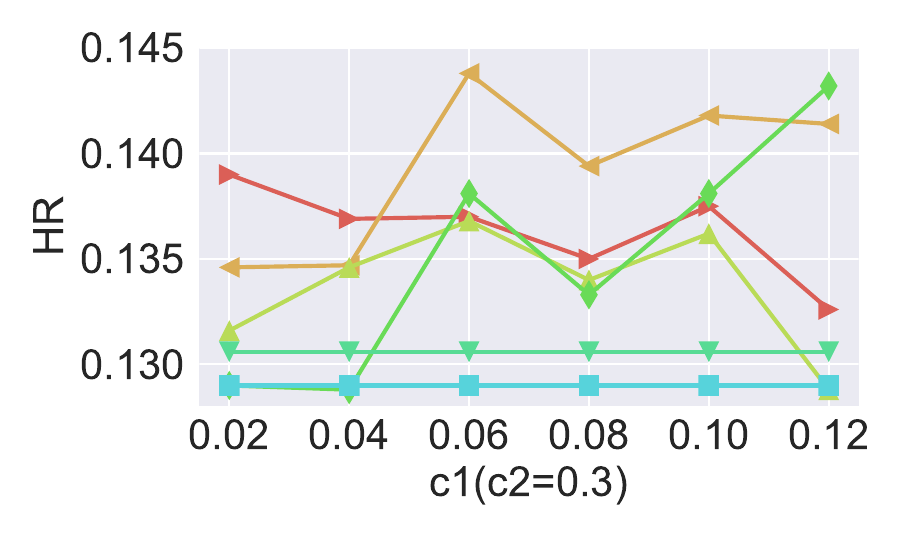}
	}
	\hspace{-.1in}
	\subfigure[Books-DR-c2]{
			\includegraphics[width=4.3cm,height=2.6cm]{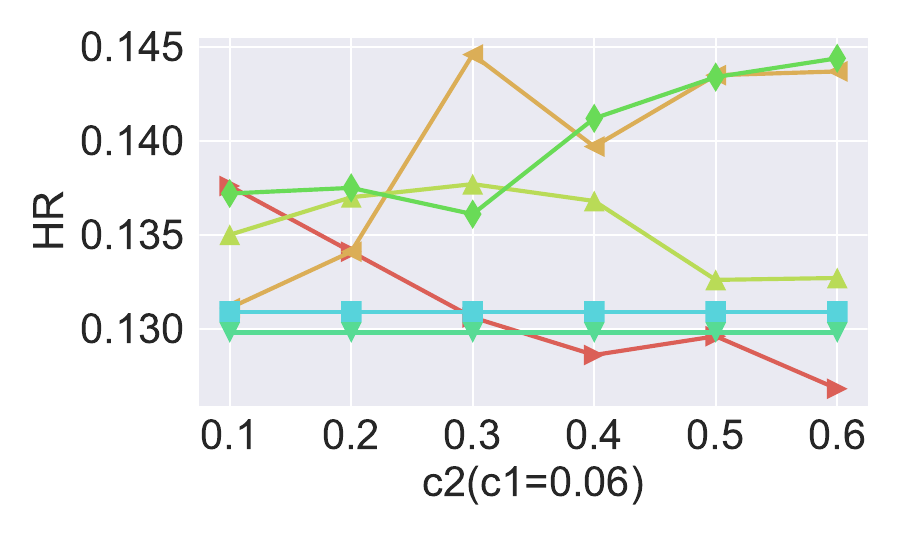}
	}
	\hspace{-.1in}
	\subfigure[Books-SA-c2]{
			\includegraphics[width=4.3cm,height=2.6cm]{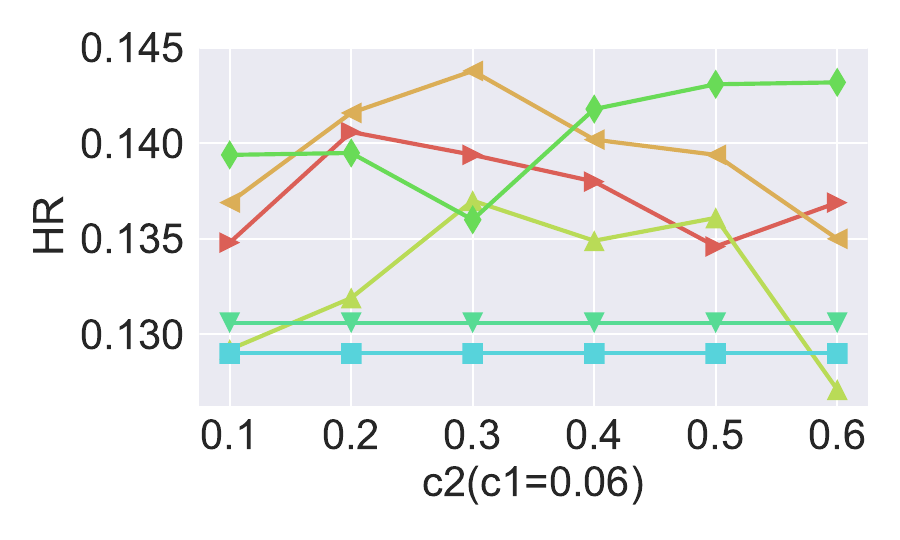}
	}
	\vspace{-.06in}
	\end{minipage}
	\begin{minipage}[b]{2\linewidth}
	\subfigure[Taobao-DR-c1]{
			\includegraphics[width=4.3cm,height=2.6cm]{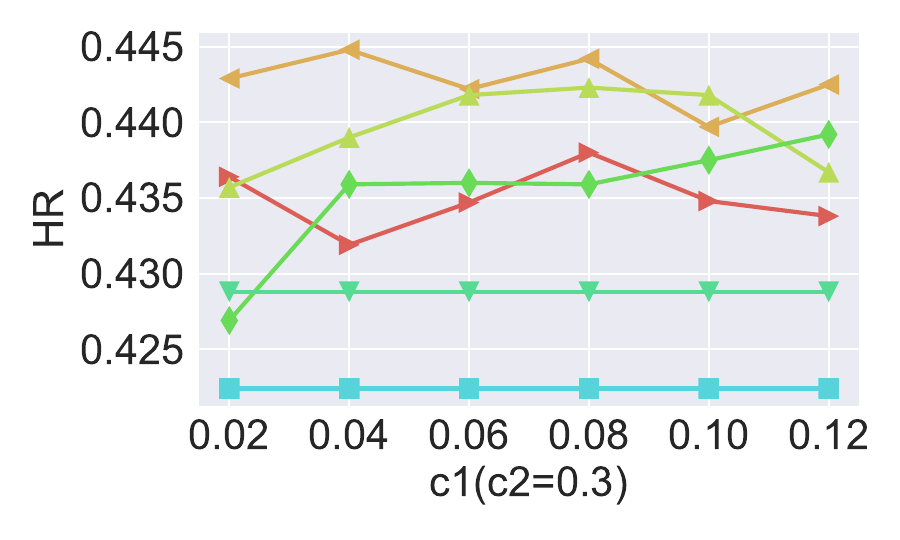}
	}
	\hspace{-.1in}
	\subfigure[Taobao-SA-c1]{
			\includegraphics[width=4.3cm,height=2.6cm]{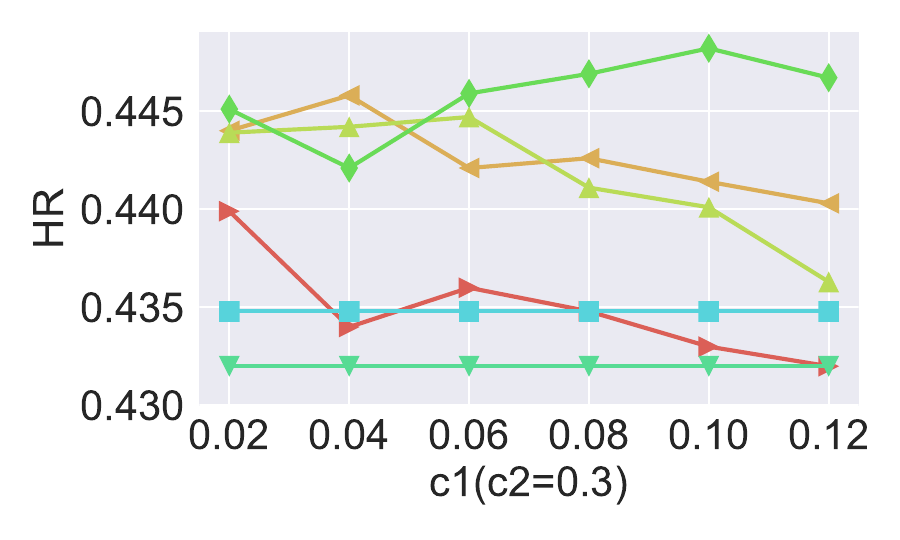}
	}
	\hspace{-.1in}
	\subfigure[Taobao-DR-c2]{
			\includegraphics[width=4.3cm,height=2.6cm]{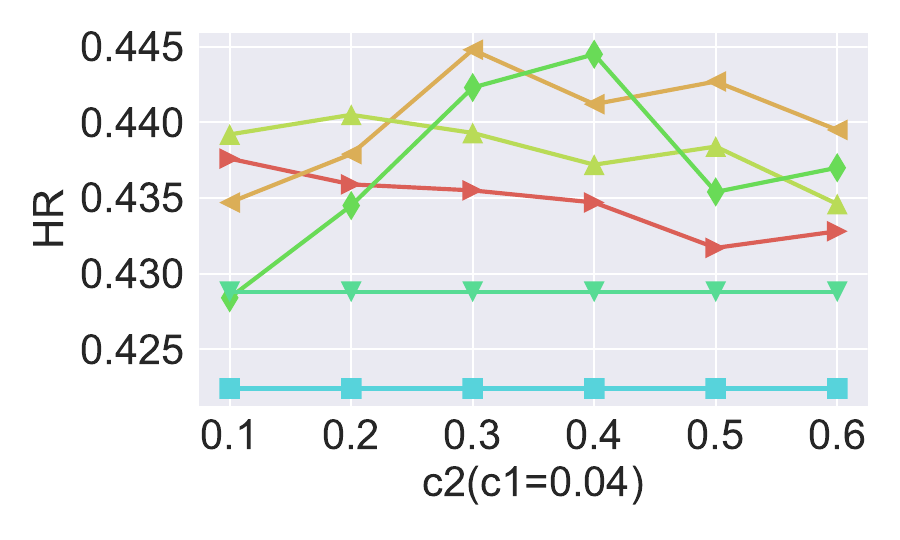}
	}
	\hspace{-.1in}
	\subfigure[Taobao-SA-c2]{
			\includegraphics[width=4.3cm,height=2.6cm]{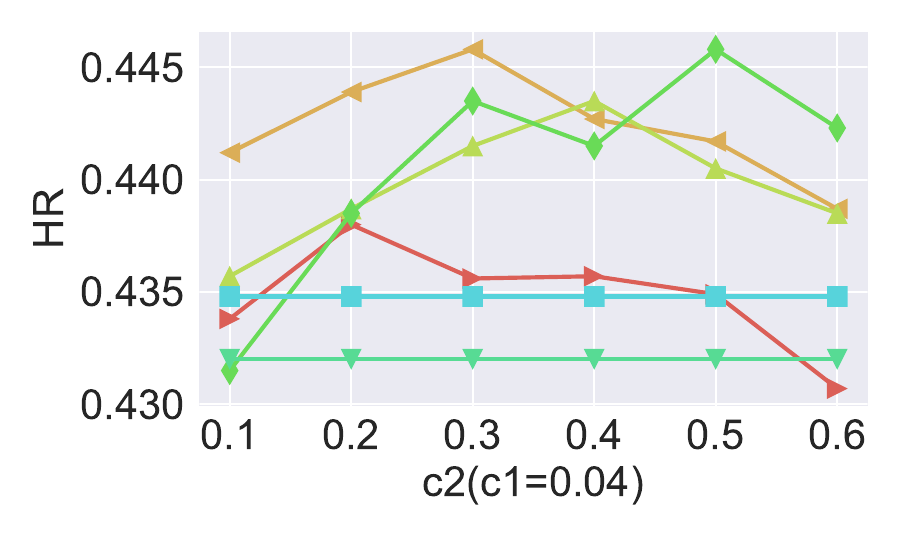}
	}
	\end{minipage}
	\vspace{-.1in}
	\caption{Performance with different $c_1$,$c_2$, initial $K$ numbers and $\delta K$ numbers on \textbf{Books} and \textbf{Taobao} datasets.}
	\label{fig:parameter}
	\vspace{-.1in}
\end{figure*}

\subsubsection{Parameter Sensitivity}  We investigate the sensitivity of the puzzlement threshold $c_1$ in new-intents detection and L2-norm threshold $c_2$ in projection-based intent trimming on \model performance together with different values of $K$ for initial intents and the different values of $\delta K $ for newly created intents. We choose the ComiRec-DR/SA as the base models and test on \textbf{Books} and \textbf{Taobao} two datasets. Similar conclusions can be drawn from other datasets and base models. 
\noindent\textbf{Hyperparameters $c_1$ and $c_2$} We first vary the value of $c_1$ in $\{0.02, 0.04, 0.06, 0.08, 0.10, 0.12\}$ and $c_2$ is set to 0.3 on both datasets. As shown in Figure~\ref{fig:parameter}, the model achieves the highest performance with moderate values of $c_1$ in most cases because too large $c_1$ prevents the creation of the new intent. 
Then we vary the value of $c_2$ in $\{0.1, 0.2, 0.3, 0.4, 0.5, 0.6\}$ and $c_1$ is set to 0.04/0.06 on \textbf{Books} and \textbf{Taobao} respectively. In Figure~\ref{fig:parameter}, the model achieves the highest performance also with moderate values of $c_2$ in most cases because too small $c_2$ prevents the trivial intents from trimming. 
\noindent\textbf{intents Number $K$ and $\delta K$} The value of $K$ and $\delta K$ are chosen from $\{(4,1),(4,3),(6,1),(6,3),(19,0),(21,0)\}$, where $K=19,$ $\delta K=0$ and $K=21,$ $\delta K=0$ equal to the settings where we create all the intent vectors in advance at the pre-training stage for \model with $K=4,\delta K=3$ and $K=6,\delta K=3$, respectively.
As shown in Figure~\ref{fig:parameter}, \model with $\delta K=3$ reports higher results than $\delta K=1$, which shows user may develop multiple intents in a new time span. \model achieves higher HR values when $K=6$ on \textbf{Taobao}, probably because users generally have more intents in \textbf{Taobao}. 
We also find that the highest performance for $K=4$ and $\delta K=1$ is achieved at smaller $c_1$ or $c_2$ probably because $K$, $\delta K$ are relatively small and we need more loose intent expansion pattern controlled by $c_1$ and $c_2$.
The performance of \model with $K=19,\delta K=0$ and $K=21,\delta K=0$ is far below the performance of \model with $K=4/6,\delta K=1/3$, which confirms the effectiveness of using intents expansion strategy and shows that the model suffer when we create too many intent vectors in advance.

\subsection{Experimental Results for \modelb}
\begin{figure}[h]
\vspace{-.1in}
\centering
	\begin{minipage}[b]{\linewidth}
	\centering
	\begin{subfigure}
   \centering
	\includegraphics[width=8.8cm]{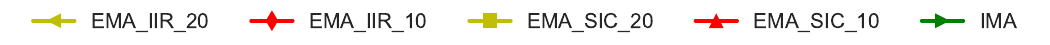}
	\vspace{-.1in}
	\end{subfigure}
	\end{minipage}
	\setcounter{subfigure}{0}	
	\centering
	\vspace{-.1in}
	\subfigure[Xlong-HR]{
			\includegraphics[width=8.8cm]{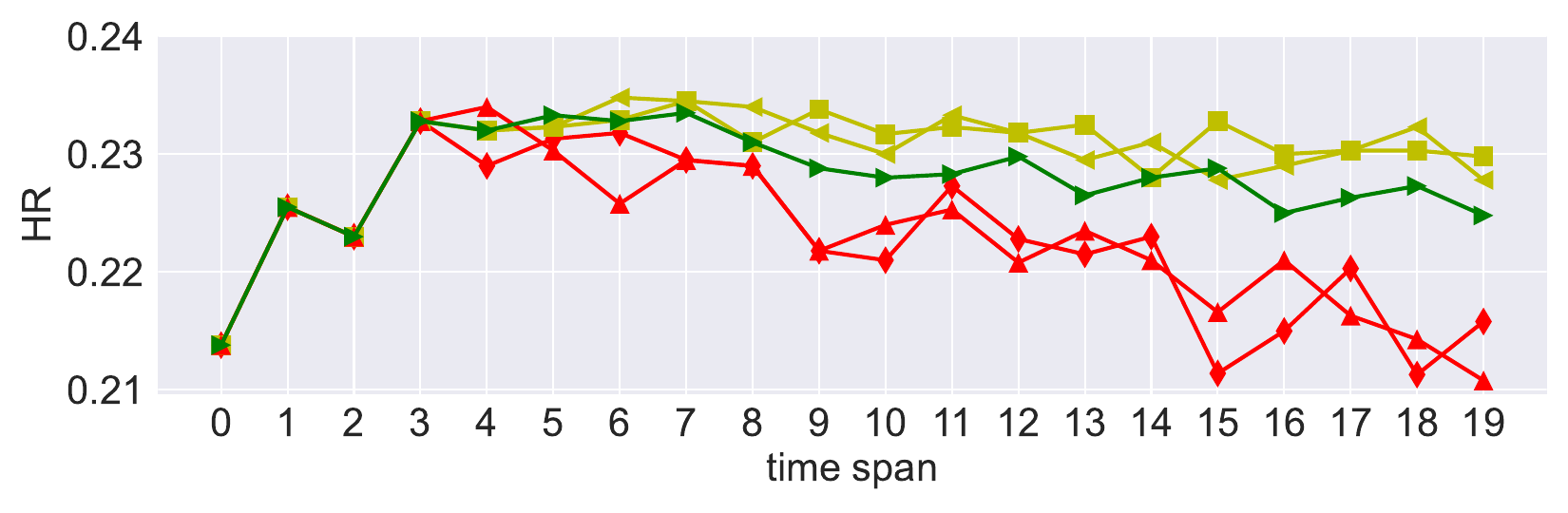}
	}
	\vspace{-.1in}
	\subfigure[Tmall-long-HR]{
			\includegraphics[width=8.8cm]{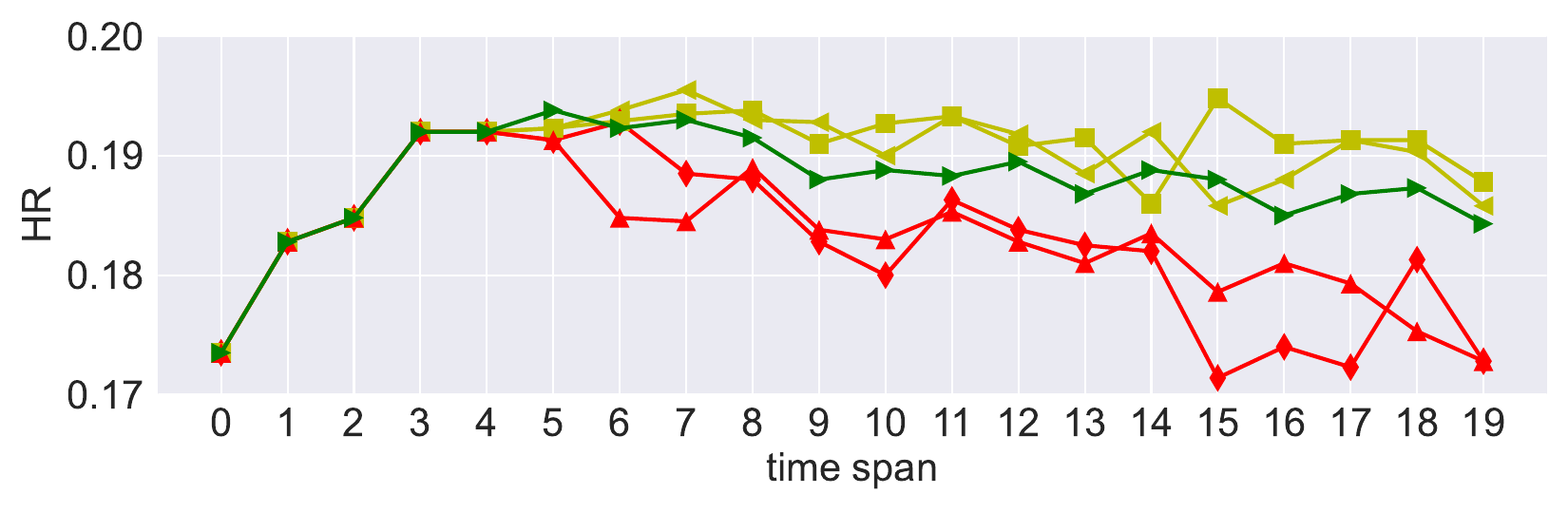}
	}
	\vspace{-.1in}
	\subfigure[\#intent of Xlong]{
			\includegraphics[width=8.8cm]{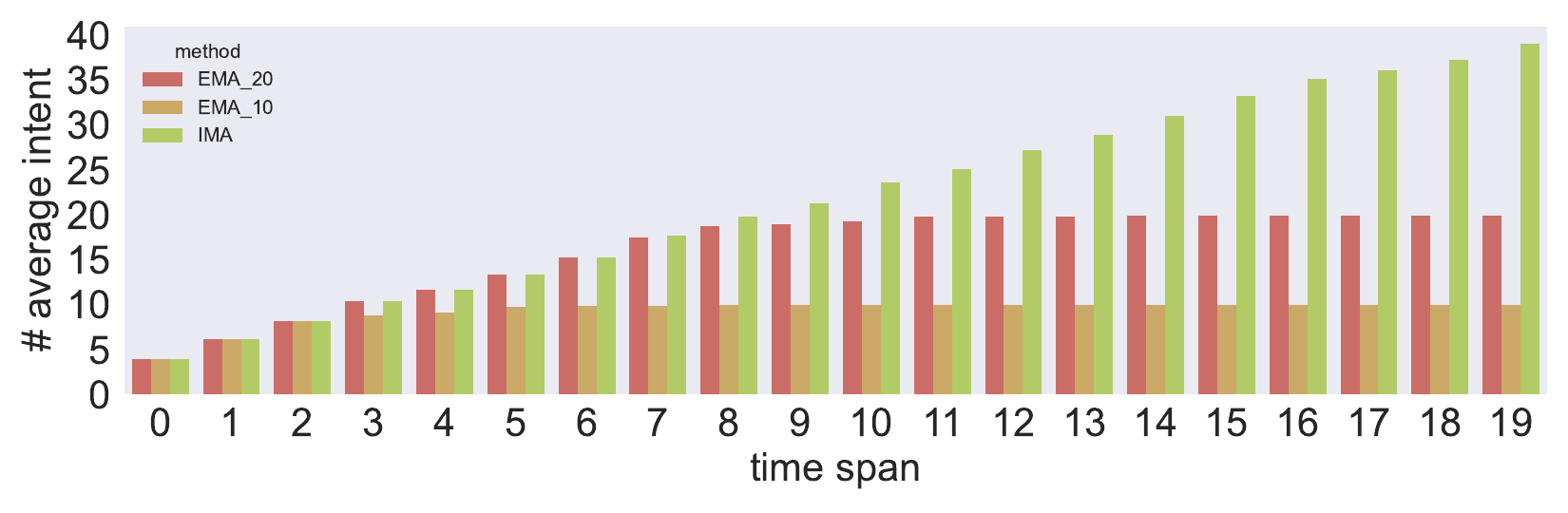}
	}
	\vspace{-.1in}
	\subfigure[\#intent of Tmall-long]{
			\includegraphics[width=8.8cm]{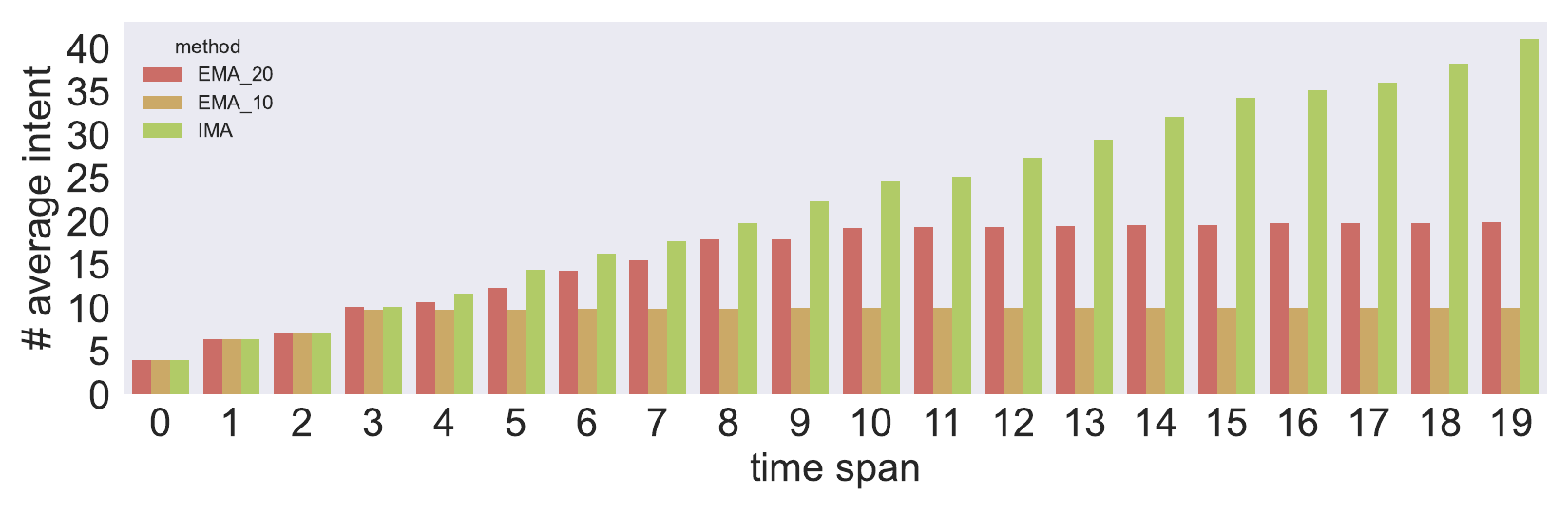}
	}
	
	\caption{Experiment result of \modelb against \model on Xlong.}
	\label{fig:xlong}
\end{figure}
In Section~\ref{sec:ema}, we upgrade the \model into \modelb which can elastically remove inactive intents and compress user intent vectors under a controllable limit. In this section, we provide extra experiments on datasets with extremely long interaction sequences, which are more common in real recommender system. The experiment results verify that \modelb can effectively decrease the intent number but achieve superior performance compared to \modelns.

\noindent\textbf{Datasets} The datasets are listed below.
\begin{itemize}[itemsep=0pt,topsep=0pt,parsep=0pt,leftmargin=10pt]
\item \textbf{Xlong}~\cite{xlong}: The user interaction sequence in Xlong averagely has 732 interactions ranged in 6 months, which is much longer than the one in Taobao and Amazon. 
\item \textbf{Tmall-long}~\cite{tmall2}: The data set contains anonymized users' shopping logs in the past 6 months before and on the "Double 11" day. The interaction sequence in Tmall-long averagely has 1077 interactions. 
\end{itemize}
We split both Xlong and Tmall-long into 20 time spans according to timestamps and adopt the same evaluation process in Section~\ref{datasets}. 

\noindent\textbf{Compared Methods} The compared methods are listed as follows. IIR and SIC refer to inactive intents removal and similar intents compression, respectively.
\begin{itemize}[itemsep=0pt,topsep=0pt,parsep=0pt,leftmargin=10pt]
\item \textbf{\modelnsb\_IIR\_20/10} we remove extra intents using IIR once a user's intent number exceeds 20/10, where 20/10 is the upper limit of intent numbers $K_{max}$. 
\item \textbf{\modelnsb\_SIC\_20/10} we activate the SIC module to compress similar intents once a user's intent number exceeds 20/10. 
\item \textbf{\modelns} The control group. 
\end{itemize}
\noindent\textbf{Experiment Results}
Fig.~\ref{fig:xlong} shows the HR and average intent numbers in each time span for \modelb and \model. 
We have the following observations. 
First, the upper limit $K_{max}$ of \modelb will influence the performance of \modelb dramatically, because user's real intent numbers heavily exceed the upper limit but the model has no enough capacity to capture them all. For instance, the average HRs in last 10 time spans drop 4.7\% and 3.8\% for \modelnsb\_IIR\_10 and \modelnsb\_SIC\_10 compared to \model on Tmall-long.
Second, the performance of \model decrease in later time spans. The reason might be that similar intents or inactive intents interfere the model inference.
Third, both \modelnsb\_IIR and \modelnsb\_SIC with $K_{max}=20$ can achieve superior performance against IMA and decrease the average intent number to 50\%, which save 27.3\% memory. Higher percentage of memory can be saved in datasets with longer sequences.

\section{Related Work}
\subsection{Multi-intent Sequential Recommendation}
In recent years, there is a growing number of works researching personalized recommendations~\cite{icde1,icde2,fpmc,hrm,GRU4REC,fossil,caser}. One species of them typically attempts to learn user representations from historical user interaction sequences, which is referred to as sequential recommendation. 
Recently, some researchers have utilized vector networks or self-attention to capture users' diverse intents to solve this problem. 
ComiRec~\cite{ComiRec} implement both capsule-network-based and self-attention-based sequence recommendation model, which optimizes the dynamic routing process in MIND and introduces a controllable method for intent selection. 
Unfortunately, existing capsule-network-based methods cannot preserve the sequential order among interactions and lack the ability to utilize temporal information for sequential recommendation. 
Note that a recent work called LimaRec~\cite{limarec} employs linear self-attention to identify relevant information from users’ interaction sequences with different intents. This work differs from ours in the sense that LimaRec focuses on incrementally updating users’ representations for online inference. LimaRec does not provide an incremental strategy for MSR model updating and still needs to fully retrain the model using the entire historical interaction sequence periodically.

\subsection{Incremental Learning for Recommendation}
Practical recommender systems 
are often desirable to retrain the model on both historical and new interaction data to capture long-term and short-term user intents. However, full retraining would be very time-consuming and incur high memory cost.
A more cost-effective way is to perform incremental learning with new interactions in each time span~\cite{clrev,mcf,rotate}. 
There are two groups of incremental methods: model-based methods and sample-based methods. The model-based methods' main idea is to add some regularization on the loss function of new tasks to protect the existing knowledge from forgetting. Most of these kinds of methods do not require existing datasets for the model to review. 
The sample-based method focuses on which part of existing datasets should be preserved and how to combine the existing and new datasets for model training. 
In the recommendation literature, several methods have been proposed for incremental model training~\cite{ICFR,oneperson}. 
For instance, MAN~\cite{man} uses a memory augmented neural model for the incremental session-based recommendation, which memorizes a subset of testing data to enrich the training datasets on the interfere stage. 
However, among all these incremental methods for recommendation, there are no specialized incremental strategies for sequential recommendation. 
Another group of works focus on incrementally updating users’ representations based on real-time interactions during online inference. Zhou et al.~\cite{sar} proposed an expansion technique on each user's intents to generate diverse recommendation results based on streaming interaction data. 
The above work need to fully retrain the model using all historical interactions periodically. Different from these works, we focus on updating both the intents number and model parameters using the newly collected interaction data during offline training.

\section{conclusion}
In this work, we show how the existing multi-intent sequential recommender system can be deployed in incremental scenarios by using fine-tuning strategies. Furthermore, we propose an incremental learning framework for multi-intent sequential recommendation named \modelns, which augments the base dynamic routing or self-attention-based MSR models with the existing-intents retainer~(EIR), new-intents detector~(NID), and projection-based intents trimmer~(PIT) to alleviate the existing-intents forgetting problem and adaptively increase the number of the intents. We furtherly upgrade the \model into \modelb which can elastically compress user intent vectors and remove inactive intents under memory space limit.
Extensive experiments verify the effectiveness of the proposed \model and \modelb on six real-world datasets, compared with the baseline methods.
For future work, we aim to explore methods for jointly modeling intents across users, rather than maintaining separate intent vectors for each individual. This approach could uncover shared patterns and preferences, leading to more integrated and efficient intent representations.

\ifCLASSOPTIONcompsoc
  \section*{Acknowledgments}
\else
  \section*{Acknowledgment}
\fi
This work is supported by the National Key Research and Development Program of China (2022YFE0200500), Shanghai Municipal Science and Technology Major Project (2021SHZDZX0102), the Tencent Wechat Rhino-Bird Focused Research Program. We thank the anonymous reviewer for his/her careful reading of our manuscript and his/her many insightful comments and suggestions.

\ifCLASSOPTIONcaptionsoff
  \newpage
\fi



%
%
%

\bibliographystyle{IEEEtran}
\bibliography{bare_jrnl_compsoc}

%
\begin{IEEEbiography}[{\includegraphics[width=1in,height=1.25in,clip,keepaspectratio]{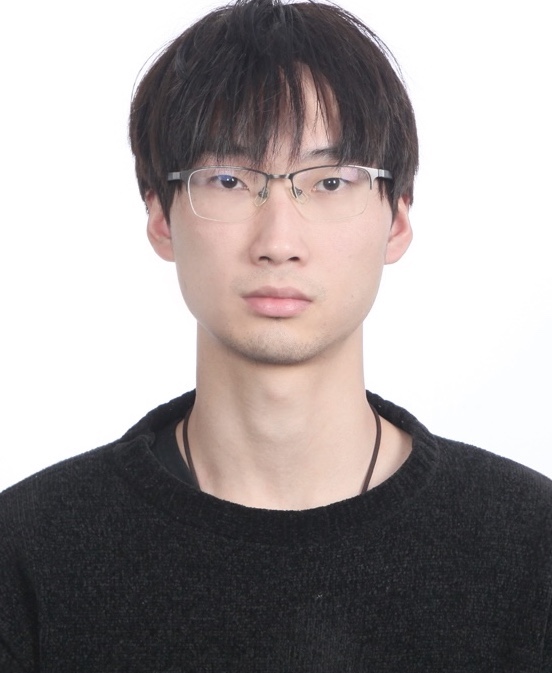}}]{Zhikai Wang}
received the bachelor degree from Shanghai Jiao Tong University, Shanghai, China, in 2019. He is a fifth year PhD candidate in the Department of Computer Science and Engineering, Shanghai Jiao Tong University. His research interests include incremental learning, recommendation system and self-supervised learning.
\end{IEEEbiography}

\begin{IEEEbiography}[{\includegraphics[width=1in,height=1.25in,clip,keepaspectratio]{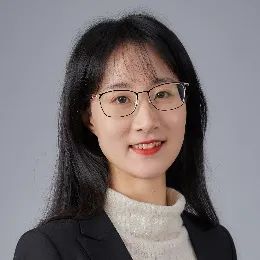}}]{Yanyan Shen}
is currently a tenured associate professor at the Department of Computer Science and Engineering, Shanghai Jiao Tong University (SJTU). She received her bachelor degree from Peking University (PKU), and obtained her doctoral degree from National University of Singapore (NUS). Her broad research interests include: databases, data mining and machine learning. She focuses on developing efficient and automated solutions to facilitate data analytics in various data-driven application domains including finance, e-commerce, etc.
\end{IEEEbiography}








\end{document}